\documentclass[]{elsarticle}

\usepackage{lineno,hyperref}
\usepackage{graphicx}
\usepackage{url}
\usepackage{color}
\usepackage{amsmath}
\usepackage{ulem}

\usepackage[caption=false]{subfig}
\modulolinenumbers[5]

\journal{JVCI} 

\newcommand{\MOD}[1]{\textcolor{black}{{#1}}}

\newcommand{\QF}{\text{QF}}
\newcommand{\Da}{\bar{\mathcal{D}}}
\newcommand{\Dna}{\hat{\mathcal{D}}}
\newcommand{\Cpixel}{\mathcal{C}_\text{pix}}
\newcommand{\Cnoise}{\mathcal{C}_\text{noise}}
\newcommand{\Chist}{\mathcal{C}_\text{hist}}

\begin{document}

\begin{frontmatter}

\title{Aligned and Non-Aligned Double JPEG Detection Using Convolutional Neural Networks}
%
\author{M. Barni$^1$, L. Bondi$^2$, N. Bonettini$^2$, P. Bestagini$^2$, A. Costanzo$^1$, \\ M. Maggini$^1$, B. Tondi$^1$, S. Tubaro$^2$}
\address{$^1$Department of Information Engineering and Mathematics, University of Siena - Italy\\ 
			  $^2$Dipartimento di Elettronica, Informazione e Bioingegneria, Politecnico di Milano - Italy} 
%
%
%

\begin{abstract}
Due to the wide diffusion of JPEG coding standard, the image forensic community has devoted significant attention to the development of double JPEG (DJPEG) compression detectors through the years. The ability of detecting whether an image has been compressed twice provides paramount information toward image authenticity assessment. Given the trend recently gained by convolutional neural networks (CNN) in many computer vision tasks, in this paper we propose to use CNNs for aligned and non-aligned double JPEG compression detection. In particular, we explore the capability of CNNs to capture DJPEG artifacts directly from images. Results show that the proposed CNN-based detectors achieve good performance even with small size images (i.e., 64x64), outperforming state-of-the-art solutions, especially in the non-aligned case. Besides, good results are also achieved in the commonly-recognized challenging case in which the first quality factor is larger than the second one.
\end{abstract}


\end{frontmatter}


\section{Introduction}\label{sec:introduction}

In the last decades, due to the wide availability of easy-to-use imaging software, diffusion of tampered content has become a widespread phenomenon. Among the techniques developed by the image forensic community to fight this trend \cite{Rocha2011, Piva2013}, great attention has been devoted to methods analyzing JPEG traces \cite{Popescu2004, chen2008machine}. Indeed, every time an image is stored (e.g., at shooting time directly on the acquisition device, or after editing with processing tools), it is usually saved in JPEG format. Therefore, manipulated content often undergoes JPEG re-compression. Because of this fact, detection of double JPEG compression has received great attention in image forensics, and presence of tampering is often revealed by looking for the artifacts left by JPEG re-compression. However, depending on whether second JPEG compression grid is aligned or not with the one adopted by the first compression, different artifacts are introduced. For this reason, these two scenarios are often analyzed separately
and are commonly referred to as aligned double JPEG (A-DJPEG) compression detection and non aligned double JPEG (NA-DJPEG) compression detection, respectively.

In many cases, manipulation takes place on limited parts of the image only. Therefore DJPEG traces are only left on a limited number of pixels. For this reason, being able to detect DJPEG on small image patches proves paramount for localization of manipulated regions in image forgery detection problems.
However, most of the techniques performing double JPEG detection in literature focus on estimating compression history of an image as a whole, whereas the localization of double compressed regions of
relatively small size (i.e., possibly tampered regions) has been often overlooked and only addressed in some works. 
In this paper we investigate the use of convolutional neural networks (CNNs) for the detection of A-DJPEG and NA-DJPEG even when working on small image patches (i.e., $64 \times 64$ pixel), which may be useful for forgery localization purpose.

\subsection{Prior Work on Double JPEG Detection and Localization}
It is well known that double JPEG compression leaves peculiar artifacts in the DCT domain, in particular, on histograms of block-DCT coefficients \cite{Popescu2004}. Accordingly, many proposed detection algorithms focus on the analysis of first order statistics of DCT coefficients. This is the case with the data-driven approach in \cite{SVMhistograms}, based on analysis of low-frequency block-DCT coefficients histograms, and \MOD{many  model-based approaches, e.g., the ones in \cite{FSDdoubleC,korus2016multi,taimori2016quantization} that rely on distribution of first (and sometimes second) significant digits (FSDs) in block-DCT coefficients and methods based on Benford-Fourier analysis \cite{CeciliaWIFS14,CeciliaInnsbruck}.} 
Data-driven detectors based on features derived from second-order statistics have also been proposed, e.g., \cite{chen2008machine}. A major drawback of many of these approaches is that they are designed to work on the whole image, i.e., to detect if an image has entirely undergone single or double JPEG compression
and they fail to correctly classify small blocks or image patches, due to the difficulty of estimating the statistics in these cases. Therefore, they are not applicable in a tampering detection scenario, when only part of the image has been manipulated.
%

Among the algorithms performing localization, Lin et al. \cite{lin2009fast} exploit double quantization (DQ) effect on DCT coefficients' histograms to produce a likelihood map reporting tampering probabilities for each $8 \times 8$ block of the image. This method has been refined in \cite{bianchi2011improved} through use of an improved probability model. However, spatial resolution considered by the authors for good detection accuracy with these methods is $256\times 256$, and performance drop significantly when smaller regions are considered. Besides, this method performs poorly when quality factor used for the first compression (i.e., $\QF1$) is significantly larger than the second one (i.e., $\QF2$). In \cite{amerini2014splicing}, localization of spliced regions is achieved by using FSD features of block-DCT coefficients and employing a support vector machine (SVM) classifier. Recently, in \cite{wang2016double}, authors proposed a novel method that relies on a one-dimensional CNN, designed to automatically learn discriminant features from DCT coefficients histograms. This approach outperforms both methods in \cite{bianchi2011improved} and \cite{amerini2014splicing},  achieving good detection performance with small sized images up to $64\times64$ pixel. However, all the above approaches exploit the peculiar traces left by aligned DJPEG compression and then fail to detect double compression in the non-aligned case.

In the NA-DJPEG scenario, several other methods for detecting double compression have been proposed, relying on ad-hoc features extracted from both pixel domain \cite{luo2007novel,chen2011detecting} and DCT domain \cite{qu2008convolutive,bianchi2012detection}. Specifically, in \cite{chen2011detecting} authors proposed a method able to detect both aligned and non-aligned re-compression. The scheme works by combining periodic artifacts in spatial and frequency domains. Specifically, a set of features is computed to measure periodicity of  blocking artifacts, which is altered when a NA-DJPEG compression occurs, and another set of features is used to measure periodicity of DCT coefficients, which is perturbed in presence of A-DJPEG. This approach for non-aligned re-compression detection is outperformed by \cite{bianchi2012detection}. Furthermore, in \cite{bianchi2012image}, Bianchi and Piva propose a forensic algorithm for tampering localization when DJPEG compression occurs, either aligned or not. The proposed scheme is as an extension of their analysis carried out in \cite{bianchi2011improved}, where a unified statistical model characterizing JPEG artifacts in the DCT domain is considered. However, similarly to \cite{bianchi2011improved} (and \cite{bianchi2012detection}), this scheme works well as long as $\QF2 > \QF1$; moreover, in order to achieve accurate detection, spatial resolutions lower than $256\times256$ pixel are not considered.

\subsection{Contribution}
Deep learning using convolutional neural networks (CNNs) \cite{Bengio2009,LeCun1998} has proved to be very powerful in many image classification problems, thus achieving considerable success in recent years also in steganalysis \cite{qian2015deep,pibre2015deep,Xu2016} and image forensics \cite{Bayar2016,Bondi2017,Chen2015}. By using CNNs, the classical machine learning paradigm of manually extracting characteristic features from the data is replaced by the possibility of learning discriminant information directly from data.

Motivated by this recent trend, the goal of this paper is to design CNN-based approaches able to classify single and double JPEG compressed images. Specifically, we are interested in working with small size images.

To the best of our knowledge, CNNs to perform double JPEG detection have been applied only in \cite{wang2016double}. In this paper, a one-dimensional CNN is designed to take as input a feature vector built by concatenating DCT histograms.
Since the network is fed with hand-crafted features (i.e., one-dimensional DCT histograms), the CNNs' capability of automatically learning from data is not addressed in that work.

In this paper, we consider the case in which the image is directly given as input to the network, thus fully exploiting self-learning capability of CNNs. Besides, our analysis is not limited to the case of aligned DJPEG compression, but we also consider the case of non-aligned double JPEG compression, in which the method proposed in \cite{wang2016double} is not meant to work. Specifically, the contributions of this paper are detailed in the following. Concerning A-DJPEG detection:
\begin{itemize}
	\item We refine the approach in \cite{wang2016double} by showing that DCT histograms can be computed using common and readily available CNN layers, and that correlation among DCT histograms can be exploited to increase classification accuracy on small $64\times 64$ images.
	\item We propose two alternative ways to perform detection based on CNNs with self-learned features directly from image pixels or noise residuals, showing the robustness of these algorithms in classifying images compressed with QFs different from those used for training.
\end{itemize}
Then, concerning NA-DJPEG compression:
\begin{itemize}
	\item We compare the proposed CNN-based detectors against state-of-the-art solutions \cite{bianchi2012detection,taimori2016quantization,korus2016multi}, showing that the CNN working on noise residuals significantly improves the performance especially on small $64\times 64$ images.
\item We confirm the robustness with respect to variations of QFs, showing that CNNs working on noise residuals are also able to correctly classify images compressed twice with the same QF.
\end{itemize}
Finally, when both A-DJPEG and NA-DJPEG are jointly considered we show that it is possible to use the same CNN-based methods to build a detector which works in the general case.

A strength of the proposed solutions, with respect to the most powerful state-of-the-art techniques (e.g., \cite{wang2016double,bianchi2012detection}), is that they are designed to work directly on the pixel values.
Therefore, our algorithms can detect
a double JPEG compression even when images are made available in bitmap or PNG format.
This indeed can be seen a simple yet effective antiforensic attack against the aforementioned methods which need to access to the information in the JPEG bitstream (e.g., to read quantization tables or quantized coefficients).

The paper is organized as it follows: in Section~\ref{sec:background} we give some basics on CNNs and discuss their usage in multimedia forensic applications. Then, in Section~\ref{sec:system} we introduce the problem of DJPEG compression detection addressed in the paper and present the CNN-based methods proposed. The experimental
methodology followed to evaluate the proposed techniques is then discussed in Section~\ref{sec:setup}. Finally, Section~\ref{sec:results} is devoted to the experimental results and the comparison with the state-of-the-art. Section~\ref{sec:conclusions} concludes the paper.

\section{Use of CNN architectures in Multimedia Forensics}\label{sec:background}

In this section we provide a fast overview on convolutional neural networks (CNNs) to highlight some of the founding concepts needed to understand the rest of the work. Particular attention is devoted to their application to Multimedia Forensics.

\subsection{Background on CNNs}


Convolutional neural networks are complex computational models that consist of a very high number of interconnected nodes. Each connection is associated to a numeric parameter that can be tuned in order to learn complex and non-linear functions \cite{Bengio2009, LeCun1998}. Network nodes are stacked into multiple layers, each one performing a simple operation on its input. With reference to the scheme depicted in Figure~\ref{fig:cnn}, some of the most common layers are the following:
\begin{figure}[t]
	\centering
	\includegraphics[width=1\linewidth]{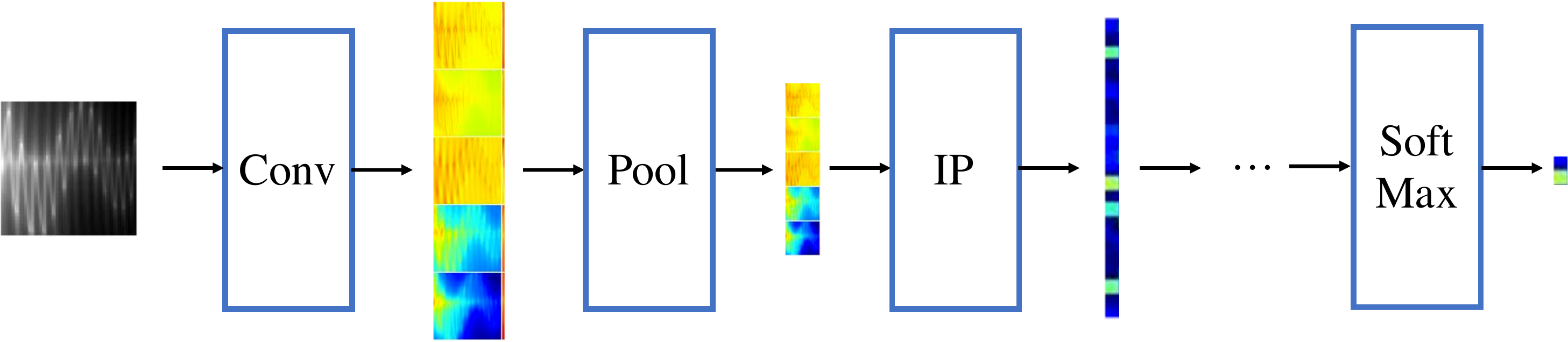}
	\caption{Example of simple CNN architecture composed by some of the principally used layers. A convolution layer (Conv) processes the input image with a series of linear filters. A pooling layer (Pool) downsamples each filtered image. Inner product (IP) linearly combines all its input data before applying a non-linear transformation. After a few other layers, SoftMax normalizes its input to sum to $1$.}
	\label{fig:cnn}
\end{figure}
\begin{itemize}
	\item \emph{Convolution}: each convolution layer is a bank of filters $h$. Given an input signal $x$, the output of each filter is the valid part of the linear convolution computed with stride $S$ (i.e., $y = \text{conv}_S(x, h)$) and it is known as feature map. The output of the layer is obtained stacking all the feature maps obtained through different filters $h$.
	\item \emph{Max-pooling}: this layer downsamples the input $x$ by sliding a small window over it and keeping the maximum value for each window position. Following the same idea, also min-pooling and average-pooling layers can be constructed.
	\item \emph{ReLU}: Rectified Linear Unit (ReLU) applies the rectification function $\max(0, x)$ to the input $x$, thus truncating negative values to zero \cite{Nair2010}. This is one of the possible way to add non-linear behavior to the network in addition to sigmoids and hyperbolic tangents among others.
	\item \emph{Inner Product}: this is a fully-connected layer that performs a set of linear combinations of all samples of the input $x$. Typically inner product layers also apply some non-linearity at the end (e.g., ReLU).
	\item \emph{SoftMax}: this normalizes the input values in the range $[0,1]$ and guarantees that they sum up to one. This is particularly useful at the end of the network in order to interpret its outputs as probability values.
\end{itemize}
By feeding the CNN with a set of labeled data (e.g., images belonging to different known categories) and minimizing a cost function at the output of the last layer, CNN weights (e.g., the values of the filters in the convolutional layers, weights for linear combinations in inner product layers, etc.) are tuned so that the CNN learns how to automatically extract distinctive features from data (e.g., image categories). This can be a great advantage with respect to the use of hand-crafted features manually defined to extract some data characteristics. Indeed, features extracted manually by following some model equations are limited by the model itself (e.g., possible linearizations, simplifications, etc.). Instead, CNNs are free to search for a better data characterization by learning it directly from the observed data, of paramount importance when defining a proper data model is an unfeasible solution. 


To train a CNN model for a specific image classification task we need: (i) to define metaparameters of the CNN, i.e., sequence of operations to be performed, number of layers,  number and shape of the filters in convolutional layers, etc; (ii) to define a proper cost function to be minimized during the training process; (iii) to prepare a (possibly large) dataset of training and test images, annotated with labels according to the specific tasks (i.e., single and double compressed JPEG images in our work).

\subsection{CNNs in Multimedia Forensics}

CNNs have been successfully used in recent years for many image recognition and classification tasks \cite{LeCun1998} and there are also many works which use CNN for applications of staganalysis, e.g. \cite{qian2015deep, pibre2015deep, Xu2016, Sedighi2017}.
However, only recently, some works have started to explore CNNs for multimedia forensic applications.

One of the first works using CNNs for multimedia forensics is \cite{Chen2015}. In this paper, the authors developed a detector for median-filtered images, whose capability of working on small $64 \times 64$ patches enabled its use also for tampering localization. In developing this algorithm, authors showed the importance of applying a pre-processing filtering step to images, in order to better expose forgery traces in a residual domain. 
The importance of working with high-pass versions of the image under analysis for forensic works was also remarked in \cite{Bayar2016}. In this paper, a CNN for forgery detection is developed, and its first convolutional layer is learn to compute high-pass-like filters. The used architecture is again very small (i.e., two convolutional and inner-product layers) as no additional depth was necessary.

A forensic task that has been better investigated with CNNs is camera model identification. To this purpose, authors of \cite{Tuama2016a} made use of a three convolutional layers network to detect the camera model used to shot a picture. In \cite{Bondi2017}, the same goal was achieved using four convolutional layers, also showing the capability of CNNs to generalize to camera models never used for training. Finally, in \cite{Bondi2017a}, the authors investigated the possibility of using up to ten convolutional layers, concluding that no additional benefit was provided by going that deep.

To the best of our knowledge, the only work based on CNNs for DJPEG detection is \cite{wang2016double}. However, in this work, the authors feed the CNN with hand-crafted features (i.e., DCT coefficients histograms) rather than letting the network learn directly from data.




\section{Double JPEG Compression Detection based on CNNs}\label{sec:system}

In this section we first introduce double JPEG detection problem, then we detail the CNN-based solutions analyzed in this work.

\subsection{Problem Formulation}
JPEG is a lossy image transform coding technique based on block-wise Discrete Cosine Transform (DCT). In a nutshell, an image is split into $8 \times 8$ non-overlapping blocks, each block is DCT transformed and quantized, then entropy coded and packed into the bitstream. Quantization is the operation causing information loss. Specifically, quantization is driven by pre-defined quantization tables scaled by a quality factor (QF). A lower QF indicates a stronger quantization, thus lower quality of the final decompressed image.
%
Double compression occurs when an image compressed with a quality factor $\QF1$ is first decompressed and then compressed again with quality factor $\QF2$. If no operations are applied between the two compression steps, $8\times 8$ JPEG blocks of the first and second compressions are perfectly aligned, thus we speak of A-DJPEG compression. Conversely, when the second compression $8\times 8$ grid is shifted with respect the previous one (e.g., due to cropping between first and second compression or to a cut and paste operation), we have a NA-DJPEG compression.
Depending on the particular scenario, both A-DJPEG and NA-DJPEG may occur.

Our goal is to build a detector which is able to classify between single compressed and double compressed images. In other words, let $H_0$ correspond to the hypothesis of single compressed image, and $H_1$ to the hypothesis of image compressed twice.  Given a $B \times B$ pixel image $I$, we want to detect whether $H_0$ or $H_1$ is verified, considering: i) only A-DJPEG case; ii) only NA-DJPEG; iii) both A-DJPEG and NA-DJPEG cases.
To solve this classification problem, we propose to use data-driven techniques based on CNNs. Specifically, starting from a standard supervised-learning pipeline, we propose three different architectures.
As it will be further explained in Section~\ref{sec:results}, the investigation of different approaches is motivated by the fact that aligned and non-aligned DJPEG compressions leave different footprints and then in principle cannot be detected in the same way
%



\subsection{Proposed Solutions}

\begin{figure}[t]
	\centering
	\includegraphics[width=.7\linewidth]{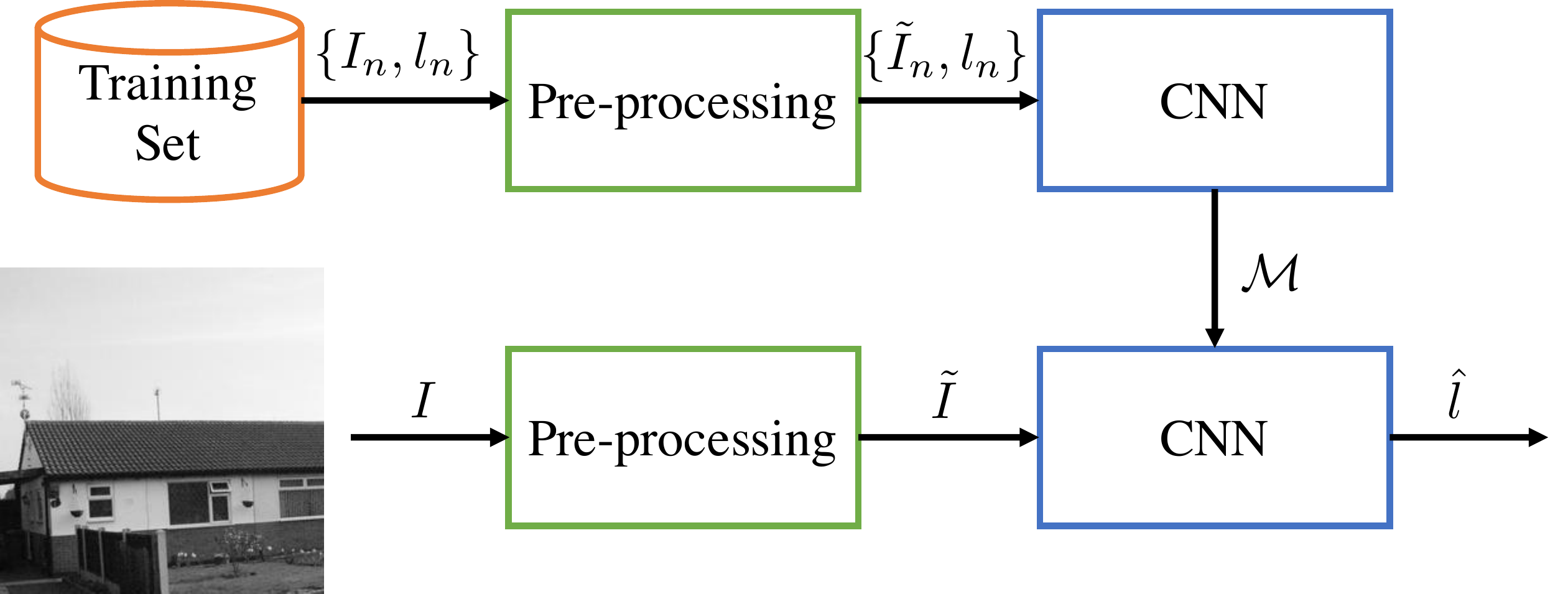}
	\caption{Pipeline common to the proposed solutions. CNN training (top) is performed using images $I_n$ labeled with $l_n$. The CNN model $\mathcal{M}$ is then used for testing (bottom) a new image $I$ and obtain the candidate label $\hat{l}$. Optional pre-processing might be applied to the images.}
	\label{fig:pipeline}
\end{figure}

The proposed methodologies follow a common pipeline depicted in Figure~\ref{fig:pipeline} composed by two steps: training and test. During training, a database of labeled images is used to learn CNN parameters for the selected architecture. Accordingly, the CNN is fed with $N$ pairs $\{I_n, l_n\}, \, n \in [1, N]$, where  $l_n = 0$ if image $I_n$ verifies $H_0$ (single compressed), $l_n = 1$ if it verifies $H_1$ (double compressed). After training, the CNN outputs the learned model $\mathcal{M}$ containing all CNN parameters (e.g., filters, fully connected weights, etc.). Optionally, a pre-processing step (e.g., denoising) can be applied to the images, in order to turn images $I_n$ into $\tilde{I}_n$. When an image $I$ is under analysis, it is fed to the trained CNN. The network outputs the probability of the image to verify whether $H_0$ is true or not. This probability (soft output) is converted to the estimated label $\hat{l}$ by thresholding (hard output). Clearly, if pre-processing is applied during training, it must be applied also during testing.

In the following we report the three investigated solutions, based on the above pipeline.

\subsubsection{CNN in the Pixel Domain}

The first investigated approach is based on the idea that properly designed CNNs should be able to automatically learn to distinguish between single and double compression by working directly  on the  image in the pixel domain. Encouraging results in this direction have been recently obtained in steganalysis field for classification of stego and cover images \cite{qian2015deep, pibre2015deep}.

In this case, $I_n$ corresponds to the JPEG images in the pixel domain (decompressed image) and\footnote{The operation is performed pixel-wise.}:
\begin{equation}
	\tilde{I}_n = I_n - \frac{1}{N}\sum_{n=1}^N I_n.
\end{equation}
The image mean subtraction is customary done before CNN training to let the network work with almost-zero-average signals.

Regarding the CNN architecture, we resort to \MOD{a slightly deeper} variation of the well-known LeNet \cite{LeCun1998a} developed for digits recognition, which has already been successfully exploited for forensic analysis \cite{Chen2015, Bayar2016, Bondi2017}. This network architecture is depicted in Figure \ref{fig:pipeline_hist} (bottom part) and input-output size of each layer are reported in Table~\ref{tab:ref_cnn}. $B \times B$ is the size of input grayscale image. \MOD{Then, three convolutional layers (i.e., Conv-1, Conv-2 and Conv-3) apply stride $1$ valid convolution with $30$ filters $5\times 5$ shaped. All of them are followed by a max-pooling layers (i.e., Pool-1, Pool-2 and Pool-3) with kernel $2\times 2$.} The first inner product layer (i.e., IP-1) reduces its input to $500$ neurons and it is followed by a ReLU non-linearity. Finally, the last fully connected layer (i.e., IP-2) reduces its input to $2$ elements, i.e., one per class. SoftMax is used at the end to normalize IP-2 output to probability values.
\begin{table}[t]
	\centering
	\caption{\MOD{Reference CNN architecture parameters. Input-output relations for each layer are reported as function of the input image size $B \times B \times 1$.}}
	\label{tab:ref_cnn}
	\resizebox{\columnwidth}{!}{
		\begin{tabular}{lccccc}
			\hline
			\textbf{\textbf{Layer}} & \textbf{\textbf{Kernel size}} & \textbf{Stride} & \textbf{Num. filters} & \textbf{Input Size}              & \textbf{Output Size}             \\ \hline
			Conv-1                    & 5$\times$5                      & 1                 & 30                      & $B \times B \times 1$              & $B$-4 $\times B$-4 $\times$ 30     \\
			Pool-1                    & 2$\times$2                      & 2                 & -                       & $B$-4 $\times B$-4 $\times$ 30     & $B$/2-2 $\times B$/2-2 $\times$ 30 \\
			Conv-2                    & 5$\times$5                      & 1                 & 30                      & $B$/2-2 $\times B$/2-2 $\times$ 30 & $B$/2-6 $\times B$/2-6 $\times$ 30 \\
			Pool-2                    & 2$\times$2                      & 2                 & -                       & $B$/2-6 $\times B$/2-6 $\times$ 30 & $B$/4-3 $\times B$/4-3 $\times$ 30 \\
			\MOD{Conv-3}                    & \MOD{5$\times$5 }                     & \MOD{1}                & \MOD{30}                      & \MOD{$B$/4-3 $\times B$/4-3 $\times$ 30} & \MOD{$B$/4-7 $\times B$/4-7 $\times$ 30} \\
			\MOD{Pool-3}                    & \MOD{2$\times$2}                      & \MOD{2}                & \MOD{-}                       & \MOD{$B$/4-7 $\times B$/4-7 $\times$ 30} & \MOD{$B$/8-3 $\times B$/8-3 $\times$ 30} \\
			IP-1                      & -                               & -                 & 500                     & $B$/8-3 $\times B$/8-3 $\times$ 30 & 500                                \\
			ReLU-1                    & -                               & -                 & -                       & 500                                & 500                                \\
			IP-2                      & -                               & -                 & 2                       & 500                                & 2                                  \\
			SoftMax                   & -                               & -                 & -                       & 2                                  & 2                                  \\ \hline
		\end{tabular}
	}
\end{table}

\subsubsection{CNN in Noise Domain}

The second solution is based on the idea that additional pre-processing, aimed at removing irrelevant information (e.g., image content), may help the CNN in its training process.

In order to expose double JPEG compression traces, we decided to rely on a denoising pre-processing operator. Then, the CNN input image $\tilde{I}_n$ corresponds to the noise residual
\begin{equation}
	\label{eq:denoise}
	\tilde{I}_n = I_n - \mathcal{F}(I_n) \;\;\;,
\end{equation}
where $\mathcal{F}(\cdot)$ is the denoising operator described in \cite{Mihcak1999}, \MOD{which relies on a spatially adaptive statistical model for the Discrete Wavelet Transform. The denoised image is predicted in the Wavelet domain by means of the minimum mean squared error (MMSE) estimation. This algorithm is} widely used in forensics for its good capability of separating image content from noise \cite{Lukas2006}.
With regard to the CNN architecture, we rely again on the one described in Table~\ref{tab:ref_cnn}.

\subsubsection{CNN Embedding  DCT Histograms}

The above solutions implicitly assume that DJPEG artifacts are exposed in the pixel domain. This is the case with non-aligned re-compressed images, which are characterized by a different behavior of blocking artifacts  with respect to single JPEG compressed one \cite{luo2007novel,chen2011detecting}. Conversely, when aligned re-compression is concerned, it is well known in the literature that peculiar traces are left in the DCT domain (specifically in the histogram DCT coefficient statistics), whereas traces left in the pixel domain are generally weaker. Therefore, our third proposed detection method relies on a CNN which automatically extract first order features from the DCT coefficients\footnote{We do not consider the case in which the image in the DCT domain is directly fed to the CNN, because, based on some preliminary experiments, we did not obtain very good performances on small images ($B = 64$).}.

Despite this approach is similar to the one proposed in \cite{wang2016double}, we would like to stress that: i) we do not make use of DCT coefficients extracted from JPEG bitstream, rather we compute DCT with a CNN layer enabling us to work with decompressed images (i.e., our method still works if double JPEG images are stored in bitmap or PNG format); ii) we exploit a 2D-convolutional CNN, rather than a 1D one as done in \cite{wang2016double}, thus capturing possible correlation among DCT coefficient histograms; iii) our solution embeds histogram computation as part of the CNN, thus enabling fast and adaptive histogram computation using one of the many available GPU frameworks for CNN; iv) by embedding histogram computation in the CNN, we are able to also optimize the choice of quantization bins, rather than fixing it manually as in any hand-crafted approach. 

\begin{figure}
	\centering
	\includegraphics[width=0.8\linewidth]{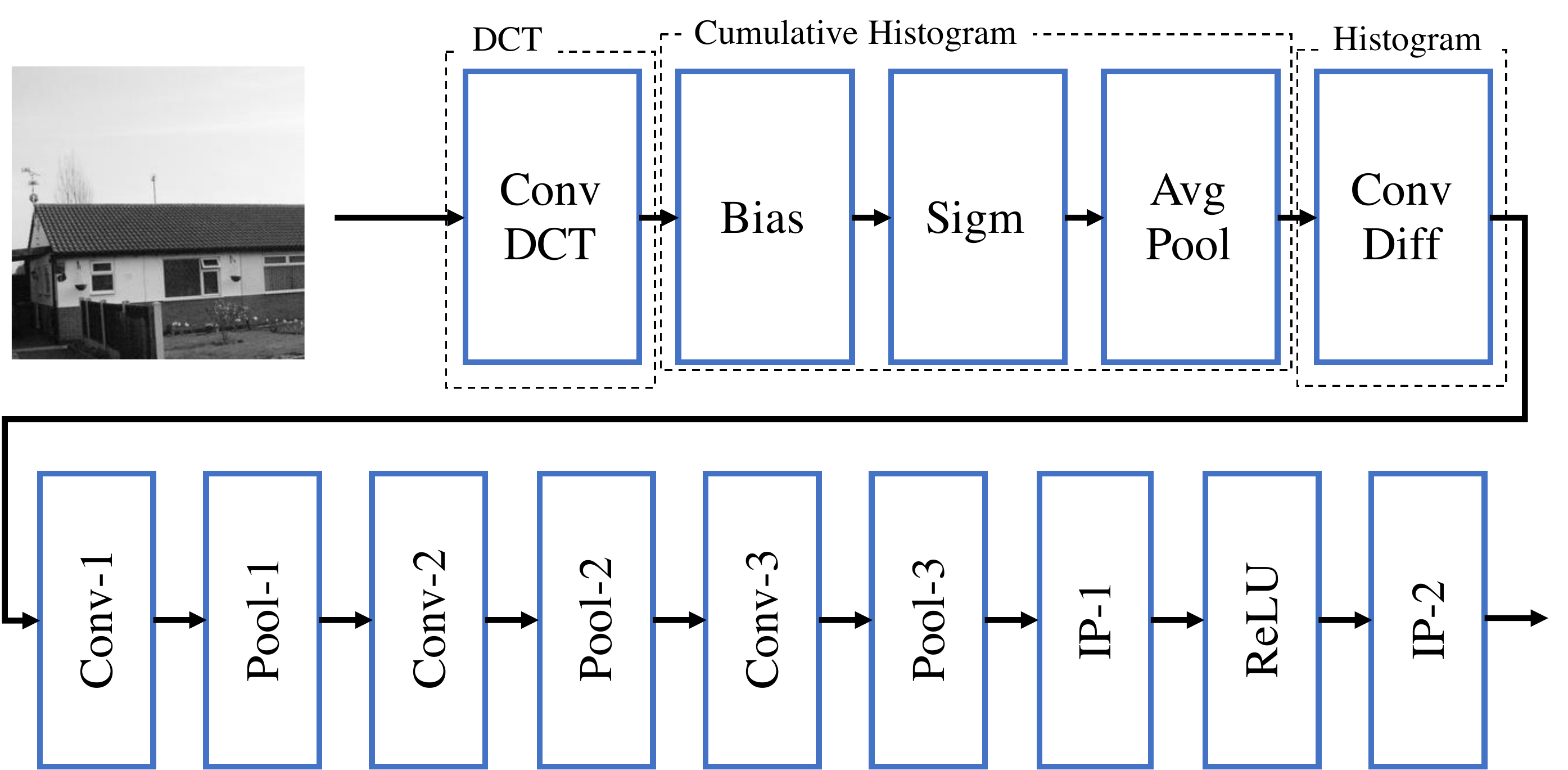}
	\caption{\MOD{Pipeline of the CNN layers used by the third proposed method. On the top, the part devoted to DCT histogram computation. On the bottom, the CNN described in Table~\ref{tab:ref_cnn}.}}
	\label{fig:pipeline_hist}
\end{figure}

Since this method does not make use of any pre-processing operation, $\tilde{I}_n = I_n$. Then, the used CNN can be thought as split into two parts as show in Figure~\ref{fig:pipeline_hist}: i) the former computes DCT coefficients histograms; ii) the latter, fed with these histogram, is the CNN described in Table~\ref{tab:ref_cnn}, whose filters in convolutional layers are $3 \times 3$ rather than $5 \times 5$.

\begin{figure}[t]
	\centering
	\subfloat[$D_{c_1, c_2}$]{\includegraphics[width=.7\linewidth]{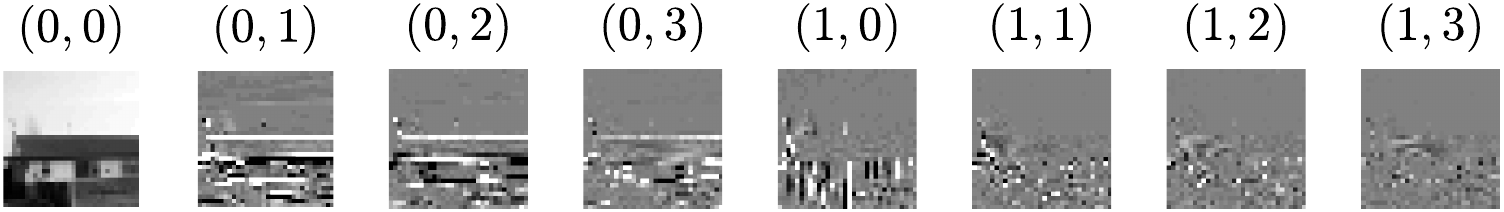}} \hfill
	\subfloat[\MOD{$D_{0, 1} - b$}]{\includegraphics[width=.7\linewidth]{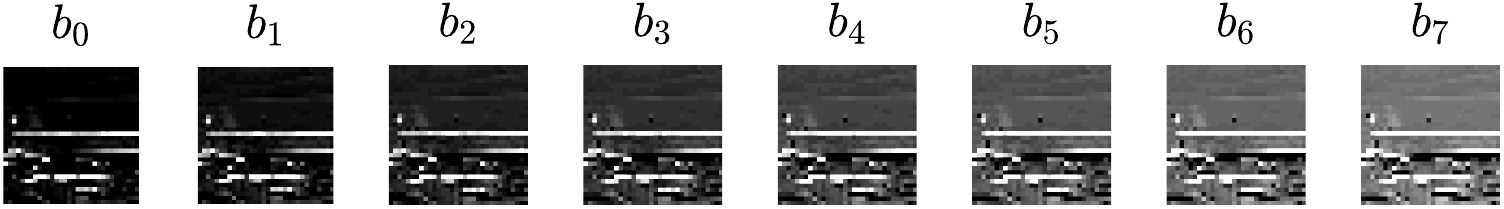}} \hfill
	\subfloat[\MOD{$\text{sigmoid} ( \gamma \cdot (D_{0, 1}(i,j) - b) )$}]{\includegraphics[width=.7\linewidth]{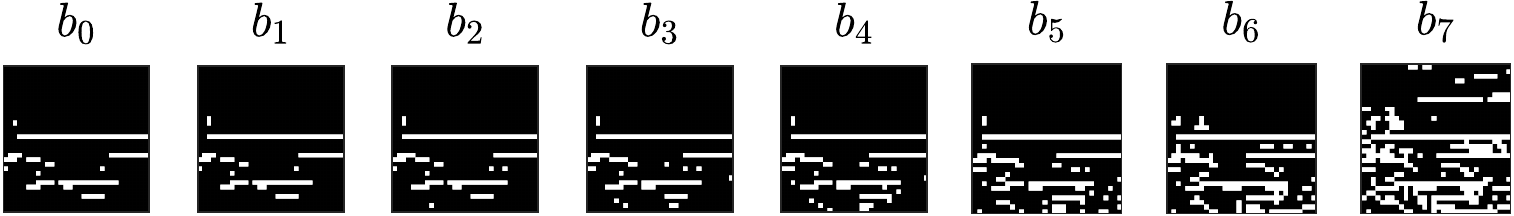}}
	\caption{Outputs of CNN layers devoted to histogram computation: (a) output of the DCT layer $D_{c_1, c_2}$ for nine different pairs $(c_1, c_2)$; (b) output of the bias layer $D_{c_1, c_2} - b$ for $(c_1, c_2)=(0,1)$ and different $b$ values; (c) output of sigmoid layer $\text{sigmoid} ( \gamma \cdot (D_{c_1, c_2}(i,j) - b) )$ for $(c_1, c_2)=(0,1)$ and different $b$ values.}
	\label{fig:dct_example}
\end{figure}

For the first part, the first step consists in obtaining the 2D DCT representation of each $8\times 8$ image block. To this purpose, let us define $D_{c_1, c_2}$ as the $\frac{B}{8} \times \frac{B}{8}$ matrix containing the DCT coefficients at frequency $(c_1, c_2)$ for each $8\times 8$ image block. This can be easily computed with a convolutional layer as
\begin{equation}
	D_{c_1, c_2} = \text{conv}_8(I, H_{c_1, c_2}),
\end{equation}
where $\text{conv}_8(\cdot, \cdot)$ computes the valid part of the 2D linear convolution using stride $8$, and $H_{c_1, c_2}$ is the DCT base at $(c_1, c_2)$ frequency.
An example of $D_{c_1, c_2}$ is reported in Figure~\ref{fig:dct_example}.

At this point, for each frequency $(c_1, c_2)$, we want to compute the histogram. To do so using common CNN layers, we first compute the cumulative histogram and then differentiate it. Specifically, to count the average number $B_{c_1, c_2}(b)$ of values in $H_{c_1, c_2}$ that are grater than a constant $b$, we resort to a series of bias, sigmoid and average-pooling layers obtaining
\MOD{
\begin{equation}
	B_{c_1, c_2}(b) = \frac{B^2}{64} \sum_{i,j \in [0,7]} \text{sigmoid}\left[ \gamma \cdot (D_{c_1, c_2}(i,j) - b) \right],
\end{equation}
}
where the bias $b$ is a constant value identifying a histogram bin boundary, $\gamma$ is a gain (i.e., $10^6$ in our experiments) used to expand the dynamic of $D_{c_1, c_2}(i,j) - b$ (i.e., to obtain very high values for $D_{c_1, c_2}(i,j) > b$ and very low values for $D_{c_1, c_2}(i,j) < b$), $\text{sigmoid}(\cdot)$ turns very high and very low input values into $0$ or $1$, and the average-pooling layer performs the sum and normalization for $\frac{B^2}{64}$. In other words, $B_{c_1, c_2}(b)$ is the $b$-th cumulative histogram bin for DCT coefficient $(c_1, c_2)$. Examples of these signals are depicted in Figure~\ref{fig:dct_example}.

The histogram for each $(c_1, c_2)$ coefficient can be obtained using a convolutional layer that computes
\begin{equation}
	Z_{c_1, c_2}(b) = \text{conv}_1(B_{c_1, c_2}, [1, -1]),
\end{equation}
where $\text{conv}$ computes 1D convolution, and the filter $[1, -1]$ acts as differentiator in the $b$-th direction. Differently from \cite{wang2016double}, we do not assume to already have access to quantized DCT coefficients. Therefore, the set of $b$ values use to construct histograms is not known and must be sought. 

Once all histograms $Z_{c_1, c_2}$ for all considered DCT frequency pairs $(c_1, c_2)$ have been computed in parallel by the CNN, they are concatenated into a 2D matrix, where each row represents a histogram bin $b$, and each column represents a frequency pair $(c_1, c_2)$. This matrix (i.e., the output of ConvDiff layer of Figure~\ref{fig:pipeline_hist}) can be considered as an image, fed as input to the CNN pipeline defined in Table~\ref{tab:ref_cnn}. 


\section{Experimental Setup}\label{sec:setup}
In this section we report all the details about experimental setup used to evaluate the proposed techniques.

\subsection{Dataset Construction}
\label{sec:setup_det}

In order to thoroughly validate the proposed solutions, we generated a set of training and test datasets of single and double compressed images at different resolutions and with different quality factors, for a total amount of more than $3$M images.
All datasets are built starting from images of RAISE database \cite{Dang-Nguyen2015}. This is a collection of more than $8\,000$ uncompressed real-world images of high resolution taken from different cameras. Images have been first converted to grayscale, then randomly cropped in order to obtain smaller resolution images used in our tests. Attention is paid to split into only one set (training or validation) all cropped portions coming from the same original image. All sets are balanced, i.e., they contain the same number of single and double JPEG images.

Training sets have been created in the following cases: i) $B = 64, 256$; ii) aligned and non-aligned DJPEG. Each set contains between $280$k and $300$k image patches. For each scenario, the image set is built as it follows: for
the first class ($H_0$), images of size $B\times B$ are single compressed with quality factor $\QF$; for the second class ($H_1$), double compressed images are built by coding $B\times B$ images first with various $\QF1$ and then with $\QF2$. For a meaningful analysis, we take $\QF = \QF2$ as done in \cite{wang2016double}.

To build double compressed images for the non-aligned case, we start from images of size $B' \times B'$ with $B' \ge B + 7$. Then, after the first compression with $QF1$, images are shifted by a random quantity $(r,c)$, $0 < r,c < 7$, and cropped to the size $B \times B$, before being compressed again with $QF2$, thus simulating grid misalignement. \MOD{In all our experiments, we consider three possible values for $\QF2$, that is $75$, $85$ and $95$, whereas $\QF1 \in \{50, 60 ,70, 80, 90\}$ for the first two $\QF2$ values and $\QF1 \in \{60, 70, 80, 90, 98\}$ for the last one.}
Table~\ref{tab:dataset_mixed} reports the breakdown of all these training datasets.
We denote with $\Da$ datasets for the aligned DJPEG case and with $\Dna$ datasets for non-aligned JPEG scenario. Superscripts indicate the adopted $\QF2$ (i.e., $75$, $85$ or $95$), whereas subscripts indicate image size (i.e., $B = 64$ or $256$).

Validation datasets have been created to evaluate: i) detection accuracy under normal working conditions, i.e., the ability of classifying test images built under the same conditions of training, and also; ii) generalization capability, that is, the ability of classifying images even when they are not perfectly compliant with the used training set. To this purpose, we generated different sets of double JPEG images with many different $(\QF1, \QF2)$ pairs and single JPEG images with the corresponding $\QF2$. Specifically, in addition to the same pairs used for training, we consider some new pairs where $\QF1$ or $\QF2$ deviates from the values used for training. Each set contains $3\,000$ single compressed images and $3\,000$ double compressed ones. As for training, validation sets are built for the case $B = 64$ and $256$, with either aligned or non-aligned DJPEG.

As commonly done to evaluate the performance with data-driven approaches, detection accuracy is measured over the same $(\QF1, \QF2)$ pairs used for training. Then, to test their generalization capability, we also measure the performance of the detectors with respect to $(\QF1, \QF2)$ pairs never used for training.

\begin{table}[t]
	\centering
	\caption{Datasets used for training. All datasets are balanced in both classes and $\QF$ pairs.}
	\label{tab:dataset_mixed}
	\scriptsize
	\vspace{1em}
	\begin{tabular}{ccccccc}
		\hline
		\textbf{Datasets}                       & \textbf{$I$ Size} & \textbf{$\QF1$} & \textbf{$\QF2$} & \textbf{Alignement} & \textbf{\# Train} & \textbf{\# Val.} \\ \hline
		$\Da^{(75)}_{256}$/$\Da^{(85)}_{256}$   & 256x256           & 50,60,70,80,90  & 75/85           & A                   & 280k              & 30k              \\
		$\Dna^{(75)}_{256}$/$\Dna^{(85)}_{256}$ & -                 & -               & -               & NA                  & -                 & -                \\
		\MOD{$\Da^{(95)}_{256}$}                     & \MOD{-}                 & \MOD{60,70,80,90,98}  & \MOD{95}              & \MOD{A}                   & \MOD{-}                 & \MOD{-}                \\
		\MOD{$\Dna^{(95)}_{256}$}                    & \MOD{-}                 & \MOD{-}               & \MOD{-}               & \MOD{NA}                  & \MOD{-}                 & \MOD{-}                \\ \hline
		$\Da^{(75)}_{64}$/$\Da^{(85)}_{64}$     & 64x64             & 50,60,70,80,90  & 75/85           & A                   & 300k              & 30k              \\
		$\Dna^{(75)}_{64}$/$\Dna^{(85)}_{64}$   & -                 & -               & -               & NA                  & -                 & -                \\
		\MOD{$\Da^{(95)}_{64}$}                      & \MOD{-}                 & \MOD{60,70,80,90,98}  & \MOD{95}              & \MOD{A}                   & \MOD{-}                 & \MOD{-}                \\
		\MOD{$\Dna^{(95)}_{64}$}                     & \MOD{-}                 & \MOD{-}               & \MOD{-}               & \MOD{NA}                  & \MOD{-}                 & \MOD{-}                \\ \hline
		&                   &                 &                 & \textbf{Tot Images} & \MOD{3 480k}            & \MOD{360k   }         
	\end{tabular}
\end{table}

\subsection{Evaluation Methodology}

In order to fairly evaluate all CNN-based considered approaches, we devised a common training-validation strategy. All CNNs have been trained using stochastic gradient descent (SGD) algorithm with batch size (i.e., number of images used for each SGD iteration) set to $128$. Momentum was set to $0.9$. Learning rate was set to $0.01$ for $64\times 64$ images and $0.001$ for $256\times 256$ images, and was progressively decreased with exponential decay at each iteration. The maximum amount of epochs (i.e., number of times the CNN sees all training data) was set to $30$ to ensure network convergence. 
As best CNN trained model, we always selected the one at the epoch with minimum validation loss in order to avoid overfitting.

The results are provided in terms of accuracy, namely the percentage of correctly classified single and double JPEG images in the validation dataset. We use notation $\Cpixel$ to refer to the CNN-based detector in the pixel domain, $\Cnoise$ for the one in the noise domain, and $\Chist$ for the case of CNN embedding DCT histogram computation. Concerning parameters of the latter, we made use of all the AC DCT frequencies. Histograms have been computed using $101$ integer bins initialized with $b\in[-50, 50]$.

\section{Results and Discussion}\label{sec:results}

In this section we evaluate the performance of the proposed detectors relying on  $\Cpixel$, $\Cnoise$ and $\Chist$, and we compare them with the state-of-the-art methods.
We first focus on the classification in the aligned double JPEG compression scenario, then we move to the case of non-aligned double JPEG compression. Finally, we provide some results in the mixed scenario of aligned and non-aligned double compression.

\subsection{Aligned Double JPEG}

\begin{figure}[t]
	\centering
	\subfloat[Impact of training set size on A-DJPEG detection accuracy using $\Cpixel$.]{\includegraphics[width=0.45\linewidth]{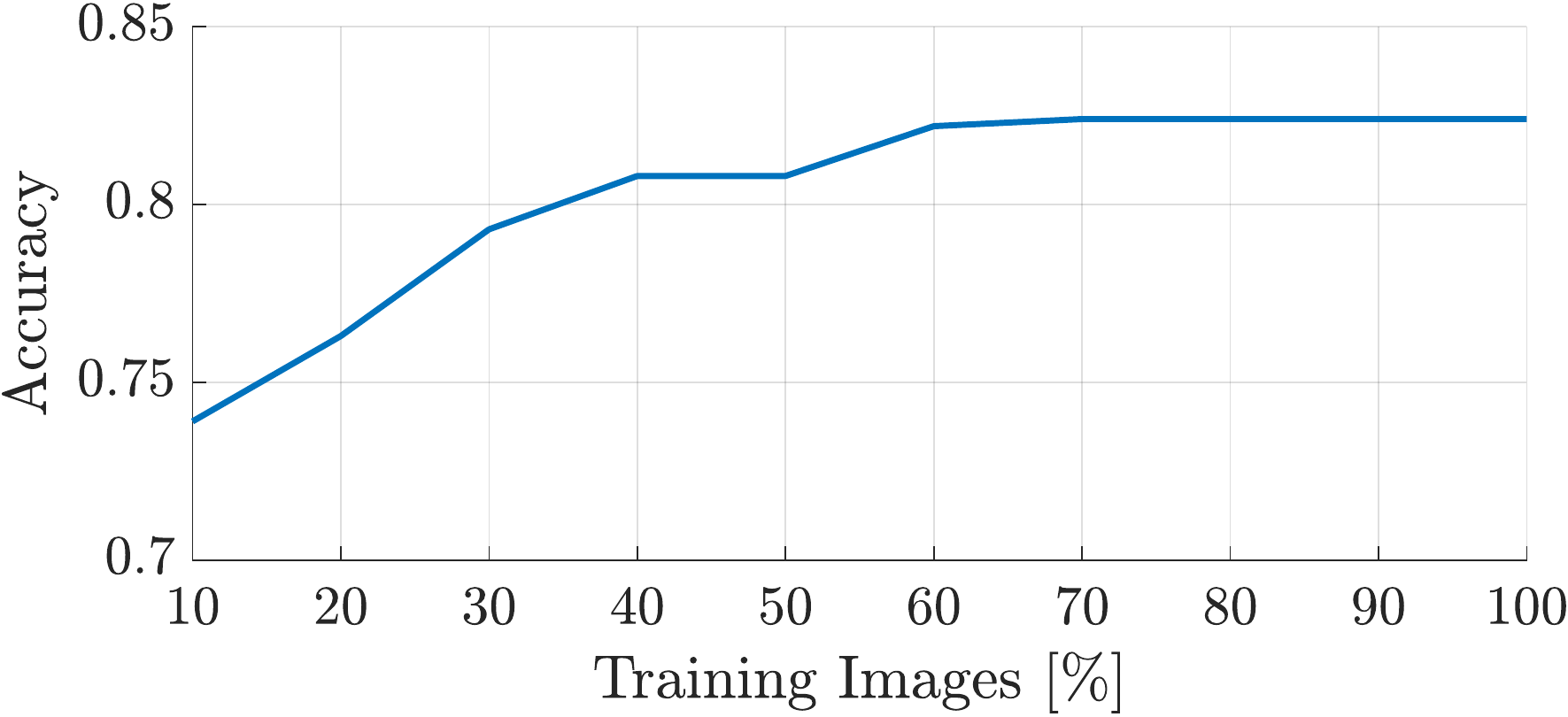}}
	\hfill
	\subfloat[\MOD{Impact of CNN depth on A-DJPEG detection accuracy using $\Cpixel$ and $\Cnoise$.}]{\includegraphics[width=0.45\linewidth]{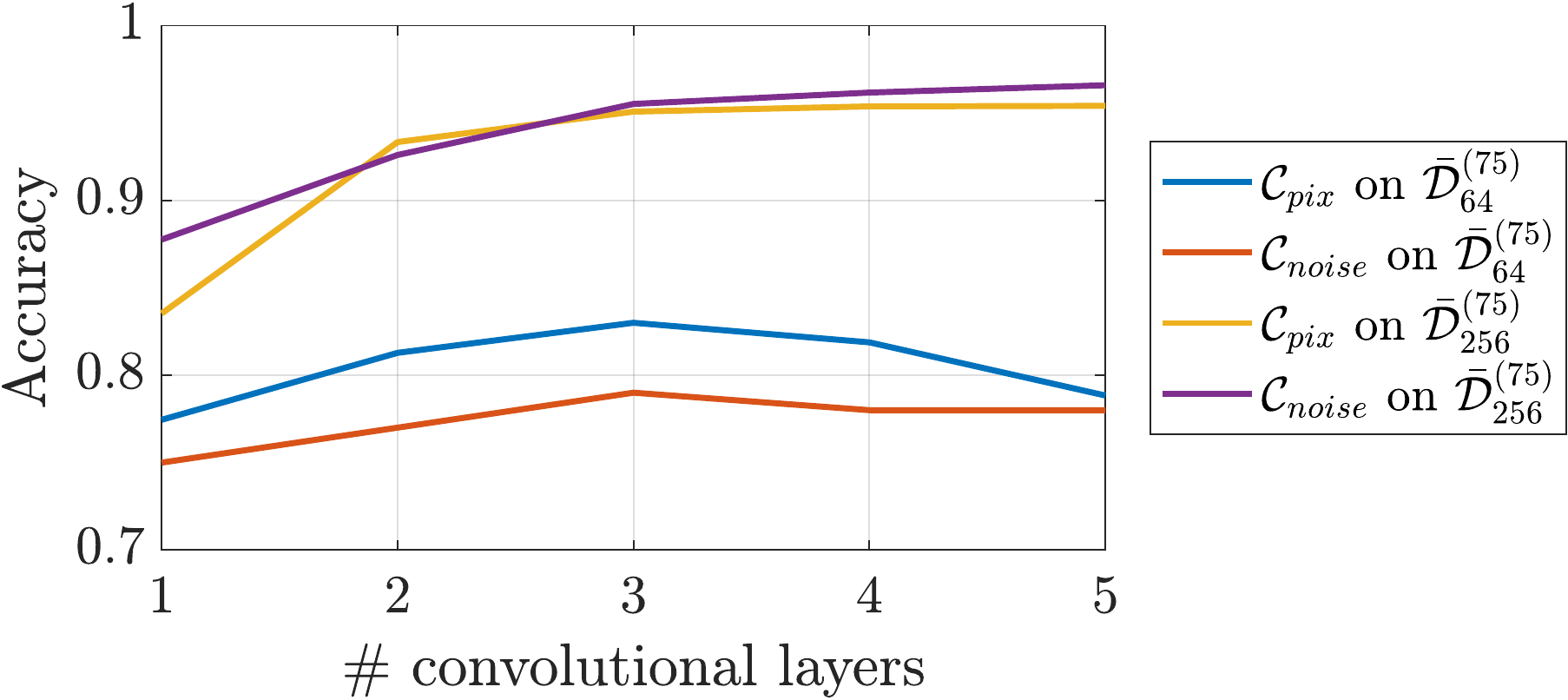}}
	
	\caption{\MOD{Impact of training set size and number of CNN layers.}}
	\label{fig:architecture_evaluation}
\end{figure}

It is well known that the performance of supervised machine learning techniques strongly depends on the amount of data used for training.
In order to assess the dependency between number of images used for training and detection accuracy in our case, Figure~\ref{fig:architecture_evaluation}(a) shows the results achieved with $\Cpixel$ in the most difficult scenario with small patches ($B=64$) and strong second quantization ($\QF=75$). To get the plot, the network is trained on different percentages of training images from $\Da_{64}^{(75)}$.
We see that, when $10\%$ of the dataset is used for training, accuracy is below $0.75$. However, when more than $70\%$ of training data is used, accuracy saturates around $0.82$.
Therefore,
using the whole training dataset, we are sure that we are not experiencing losses due to insufficient amount of training data.\footnote{It is worth pointing that the other proposed solutions, i.e., $\Cnoise$ and $\Chist$, usually need less training images to converge.}

\MOD{In order to assess the effect of CNN architecture deepness, we trained five CNNs with increasing number of Conv-Pool layer pairs on a subset of the whole dataset. Results reported in Figure~\ref{fig:architecture_evaluation}(b) show how the selected architecture almost saturates the achievable performance in terms of accuracy.}

To assess the performance of the proposed approaches for aligned double JPEG detection, \MOD{we compare them to the state-of-the-art techniques in \cite{wang2016double},\cite{korus2016multi} and \cite{taimori2016quantization}, denoted respectively as WZ, KH and TR in plot legends. We select \cite{wang2016double} as one of the baseline for two reasons: i) it is shown to outperform previously existing state-of-the-art detectors, e.g., \cite{bianchi2011improved, amerini2014splicing, FSDdoubleC, SVMhistograms}; ii) to the best of our knowledge, it is the only method based on CNNs, thus being a natural yardstick for our methods.}

\begin{figure}
	\centering
	\subfloat[\MOD{Train on $\Da_{256}^{(75)}$}]{\includegraphics[height=2.2cm, trim= 0 0 3cm 0, clip=true]{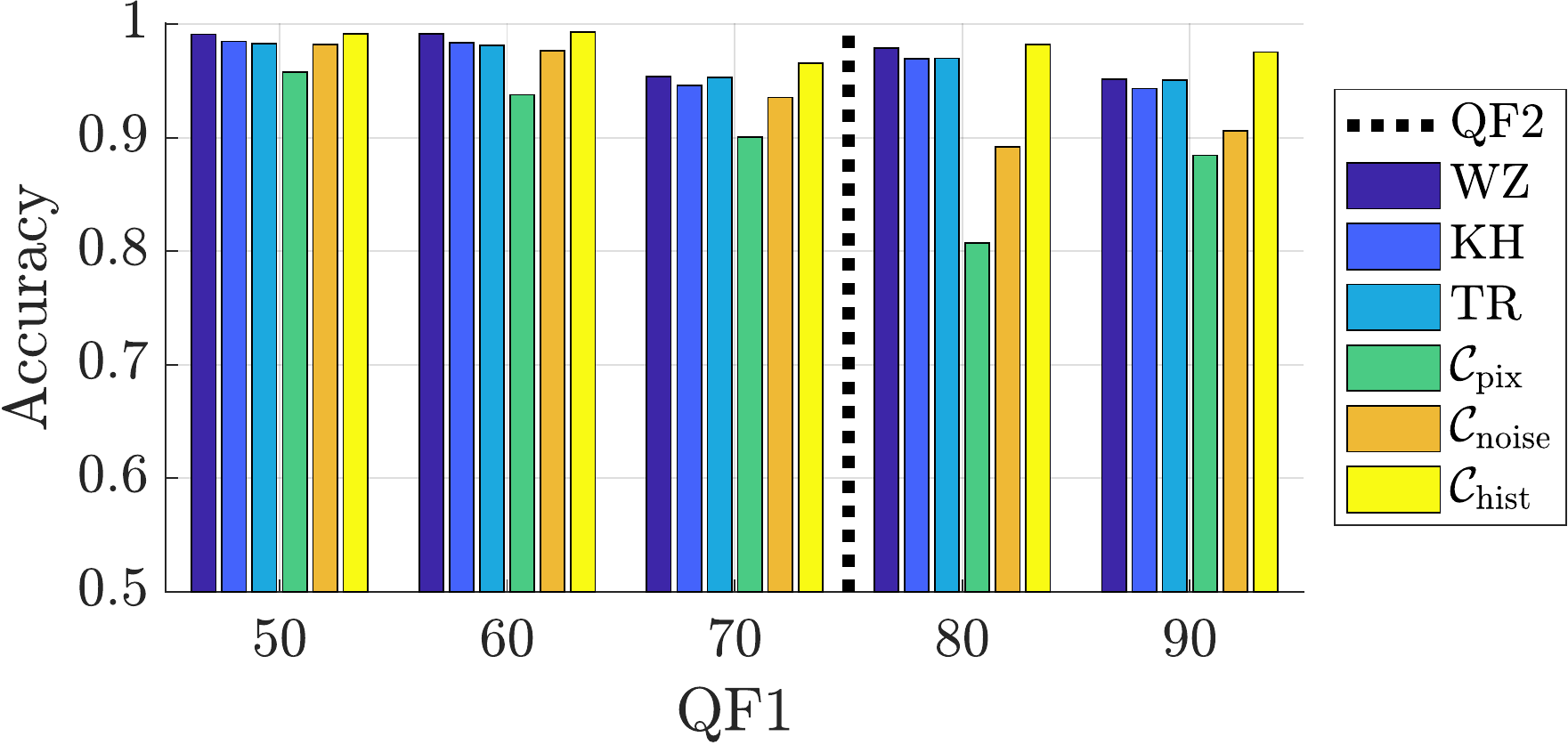}} \hfill
	\subfloat[\MOD{Train on $\Da_{256}^{(85)}$}]{\includegraphics[height=2.2cm, trim= 0.8cm 0 3cm 0, clip=true]{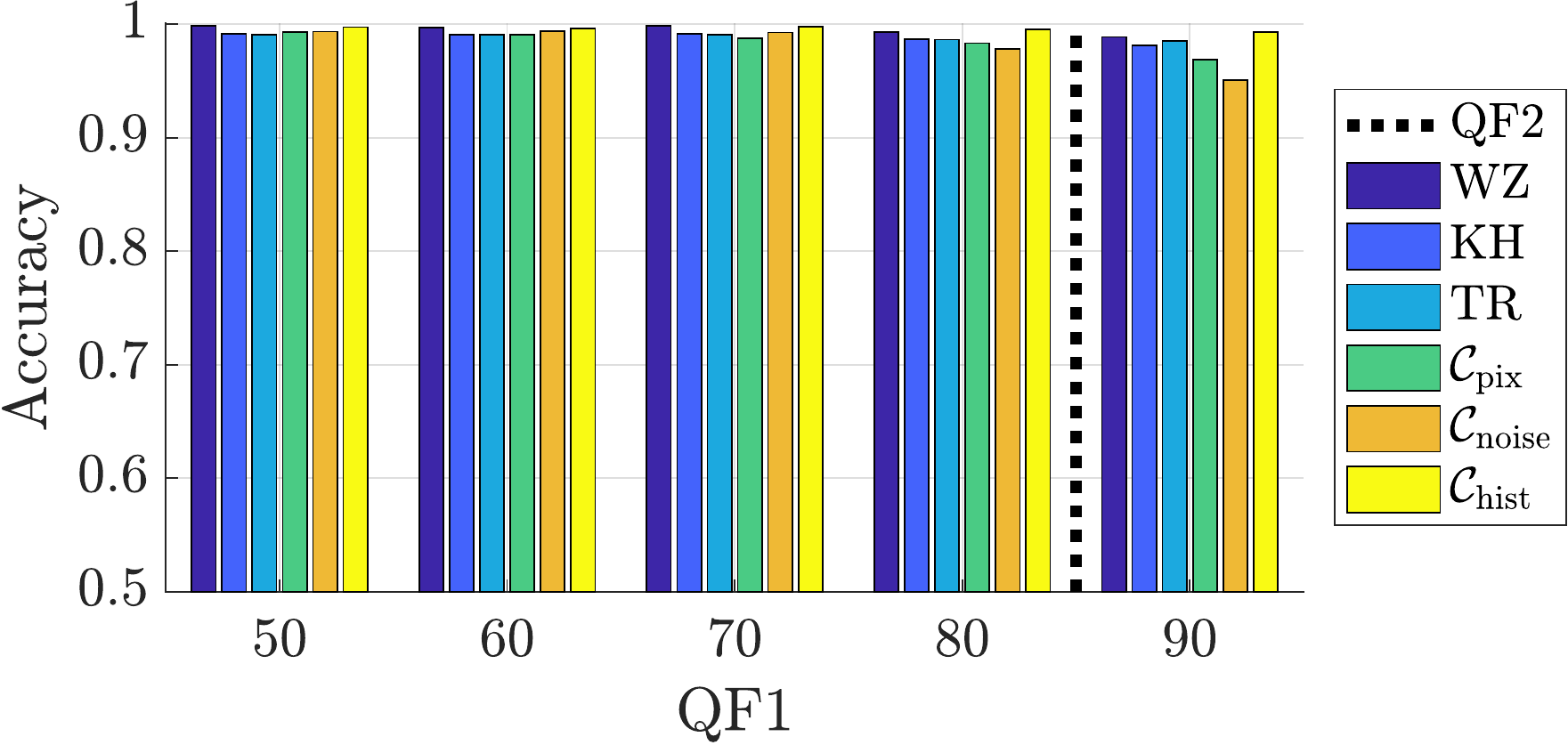}}
	\hfill
	\subfloat[\MOD{Train on $\Da_{256}^{(95)}$}]{\includegraphics[height=2.2cm, trim= 0.8cm 0 0 0, clip=true]{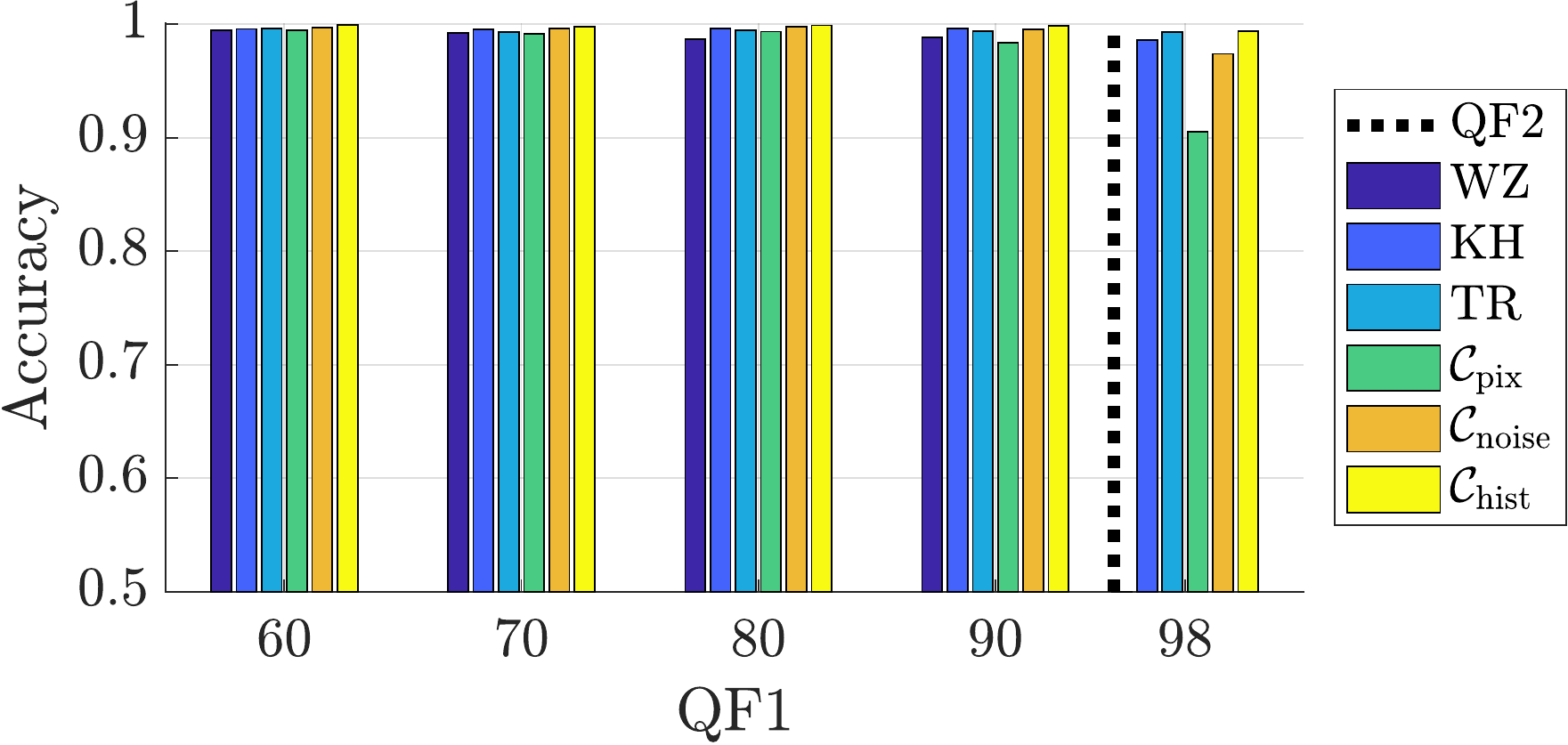}}
	
	\subfloat[\MOD{Train on $\Da_{64}^{(75)}$}]{\includegraphics[height=2.2cm, trim= 0 0 3cm 0, clip=true]{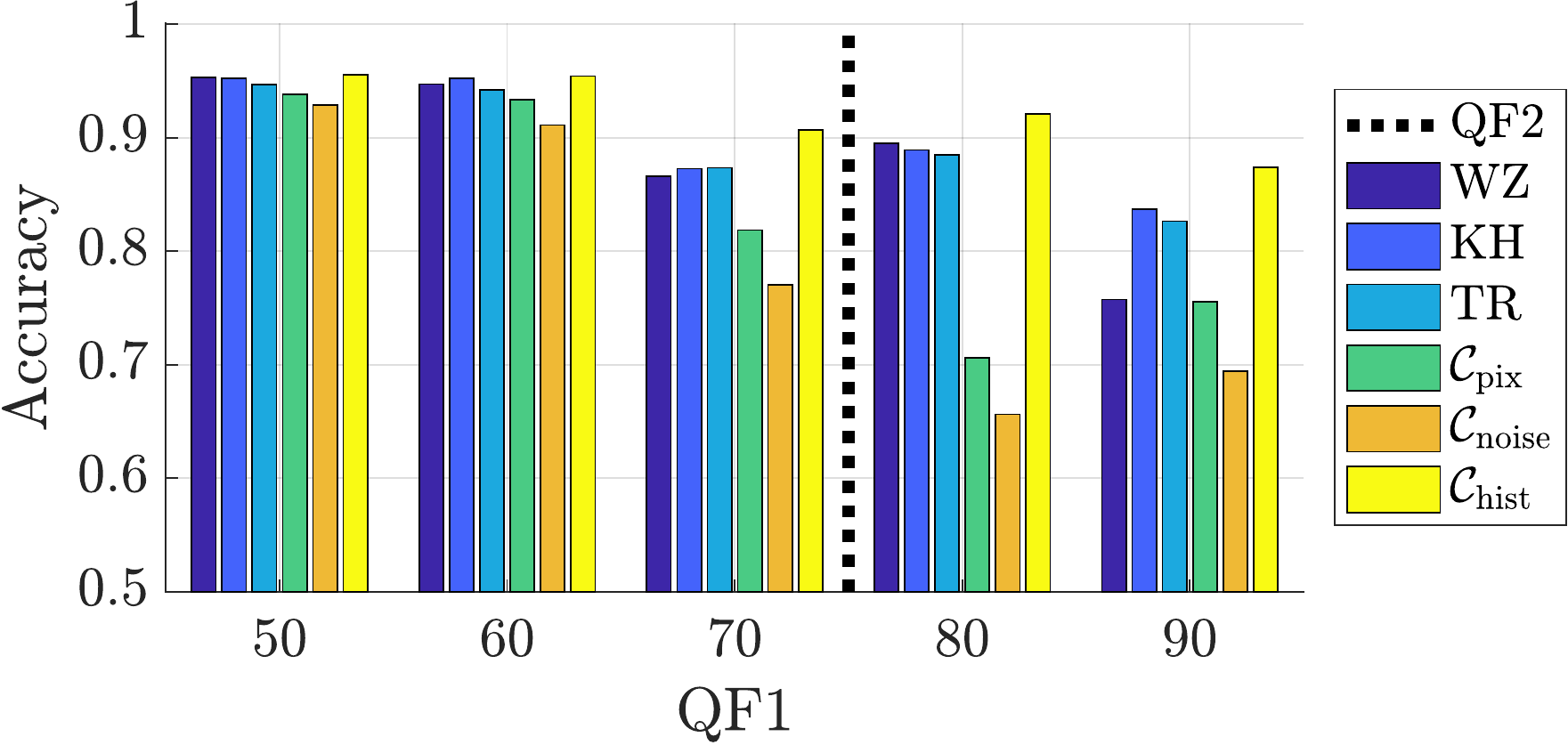}} \hfill
	\subfloat[\MOD{Train on $\Da_{64}^{(85)}$}]{\includegraphics[height=2.2cm, trim= 0.8cm 0 3cm 0, clip=true]{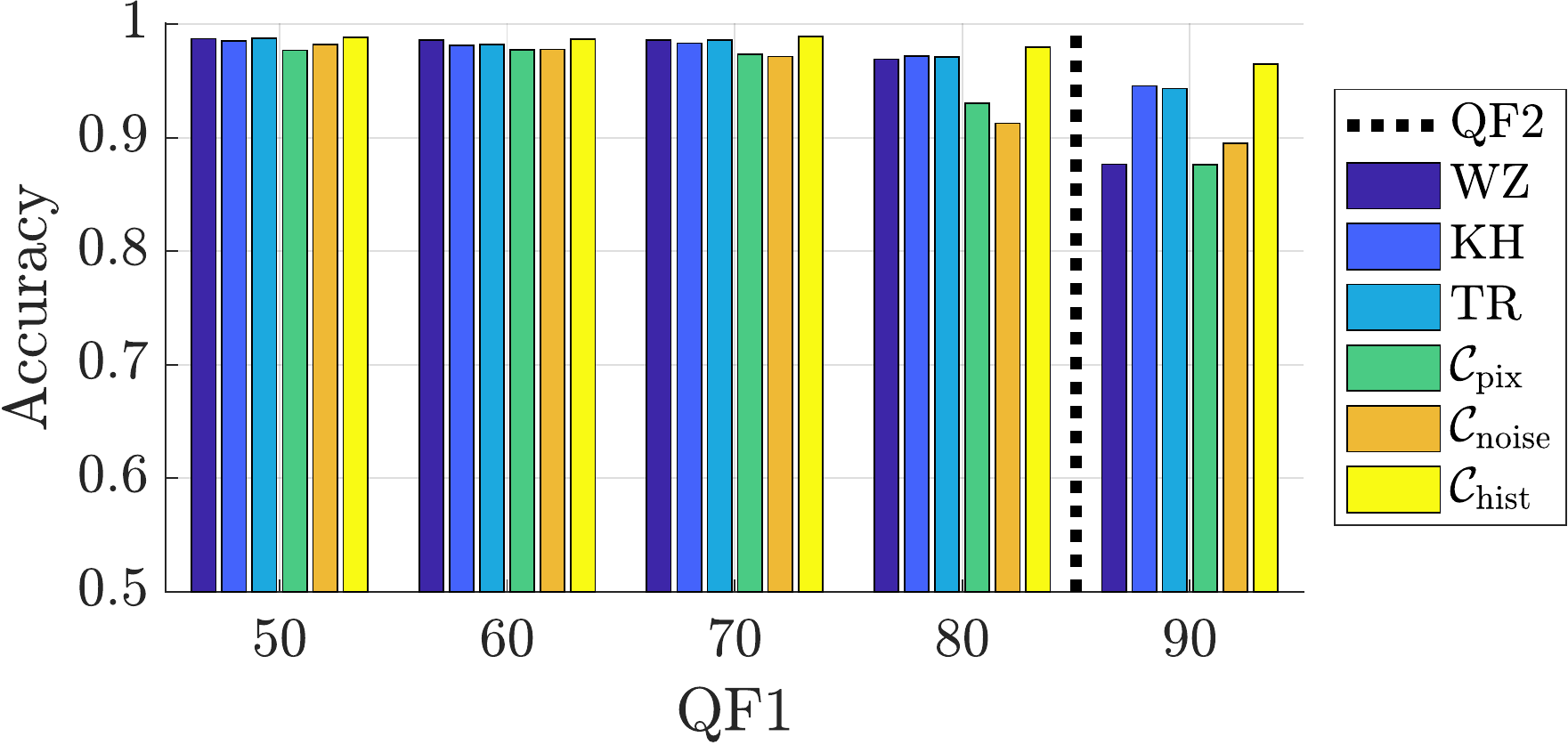}}
	\hfill
	\subfloat[\MOD{Train on $\Da_{64}^{(95)}$}]{\includegraphics[height=2.2cm, trim= 0.8cm 0 0 0, clip=true]{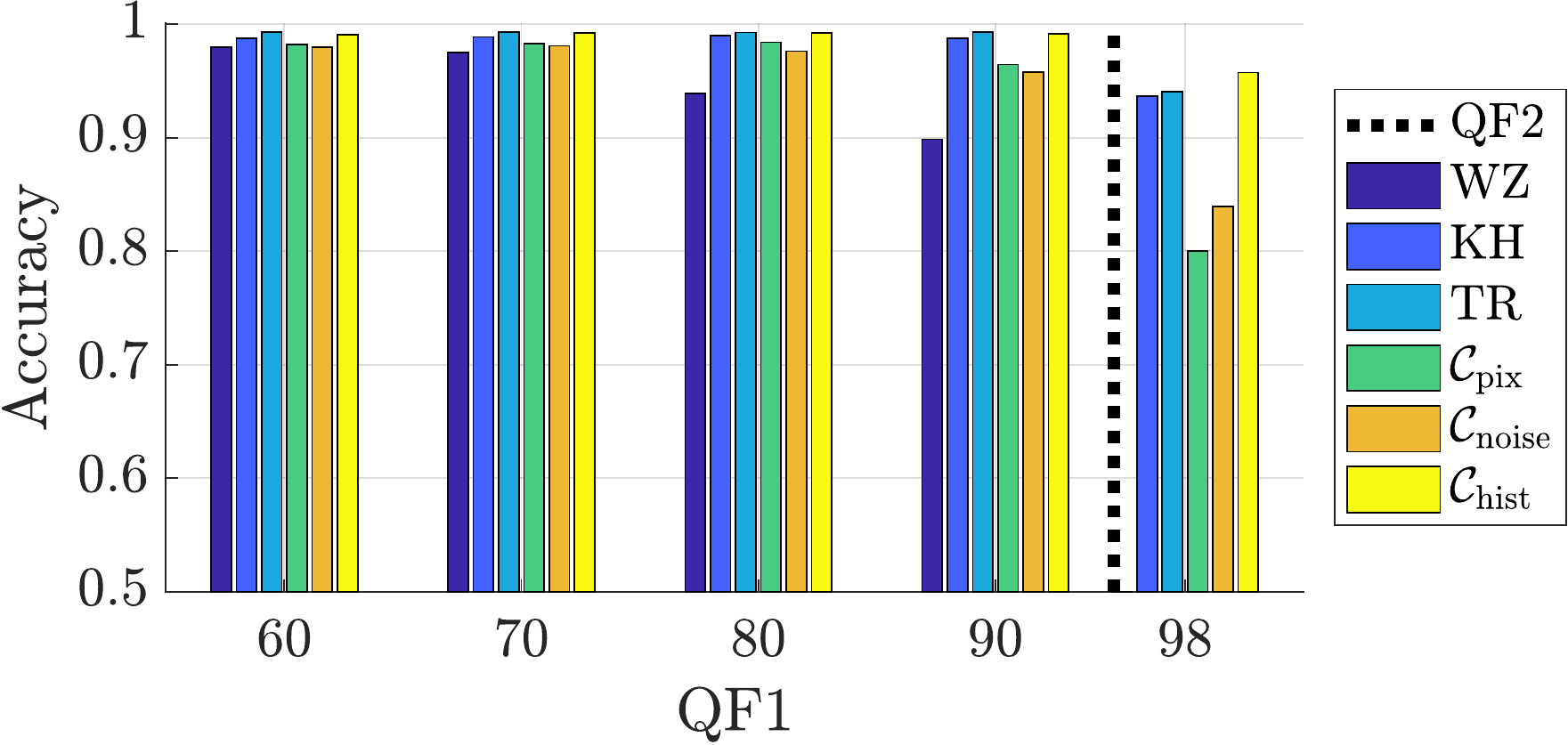}}
	\caption{\MOD{Aligned DJPEG compression detection accuracy against baselines WZ \cite{wang2016double}, KH \cite{korus2016multi}, and TR\cite{taimori2016quantization}. Dashed black line indicates the considered $\QF2$.}}
	\label{fig:acc_aligned}
\end{figure}

\MOD{Figure~\ref{fig:acc_aligned} reports results obtained training all proposed CNNs in the various cases, i.e., on the datasets $\Da_{256}^{(75)}$, $\Da_{256}^{(85)}$, $\Da_{256}^{(95)}$, $\Da_{64}^{ (75)}$, $\Da_{64}^{(85)}$ and $\Da_{64}^{(95)}$.}
Results for $B=256$ show that the proposed $\Chist$ architecture achieves equal or better performance with respect to all baseline methods. This is due to the fact that hand-crafted features exploited in  \cite{wang2016double} are very distinctive, especially when large images are concerned.

With small patches ($B=64$) all algorithms suffer when $\QF2\cong \QF1$ \MOD{(this case is addressed in the literature by specific methods tailored for the purpose, e.g., \cite{Yang_sameQm})} and $\QF2<\QF1$, as a  stronger second compression tends to mask artifacts left by the first one. However, on $64\times 64$ patches, $\Chist$ is the one with the best performance and always outperforms state-of-the-art methods on average.
Concerning the proposed methods, $\Chist$ always outperforms $\Cpixel$ and $\Cnoise$. This is also expected, as aligned DJPEG traces are better exposed in the DCT domain, rather than the pixel domain.
%
Nonetheless, a part when $\QF1$ and $\QF2$ are very close, also $\Cpixel$ and $\Cnoise$ allow to achieve accuracy greater than $0.70$ on small images.

\begin{figure}
	\centering
		\includegraphics[width=0.9\linewidth]{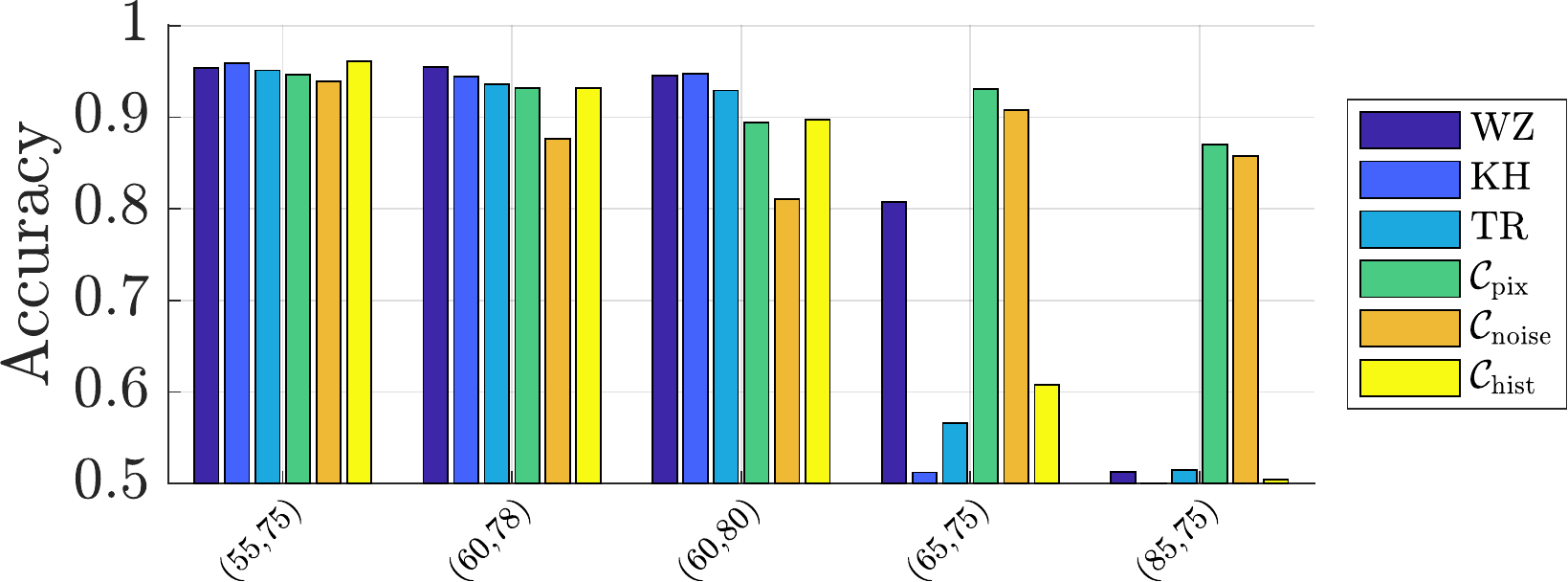}
	\caption{\MOD{Sensitivity analysis for aligned DJPEG compression detection when $\text{QF2}=75$. Image size is $64\times 64$.}}
	\label{fig:acc_sens_worst}
\end{figure}

Regarding generalization capability,
Figure~\ref{fig:acc_sens_worst} shows the accuracy achieved by all CNNs trained on the most difficult scenario
with $\QF = 75$ and small images ($B= 64$).
The methods based on DCT histograms or Benford law (i.e., $\Chist$ and baselines WZ, KH, TR) suffer
to recognize aligned DJPEG for values of $\QF1$ different from those used during training when they are close to $\QF2$, and completely fail when these $\QF1$s are larger than $\QF2$.
Contrarily,
the methods relying on pixel analysis (i.e., $\Cpixel$ and $\Cnoise$) show greater robustness to changes in $(\QF1, \QF2)$.
To further explore this fact,
Table~\ref{tab:sensitivity_aligned}(a) shows the behavior of $\Cnoise$ trained on $\Da_{64}^{(75)}$ and $\Da_{256}^{(75)}$ and tested on images with several different $(\QF1, \QF2)$ pairs.
(similar results hold for $\Cpixel$).
Similarly,  Table~\ref{tab:sensitivity_aligned}(b) reports the accuracy results with $\Cnoise$ trained on $\Da_{64}^{(85)}$ and $\Da_\text{256}^{ (85)}$. 
We notice that, by varying $\QF1$, results are perfectly in line with those achieved with matched $\QF$ pairs. Good results are also obtained with different $\QF2$s,  a part for the case of much higher $\QF2$.
\MOD{
This behavior is not surprising,
since compression with high $\QF2$ leaves few traces on images compressed at lower quality, hence detecting a DJPEG compression in these cases is hard when such examples are not included in the training set.
%
}
%


To conclude the analysis of this section,
although on one side CNNs based on a strong hand-crafted modeling assumption (as baseline \cite{wang2016double} and $\Chist$) allow to achieve the best accuracies, the ones based on the analysis of the pixel image (i.e., $\Cpixel$ and $\Cnoise$) prove to be more robust to perturbations of $\QF1$ and $\QF2$ with respect to the values used for training, which is paramount every time the algorithm works in the wild.
\begin{table}[]
	\centering
	\caption{Sensitivity of $\Cnoise$ to variations of $\QF1$ and $\QF2$ for aligned DJPEG detection.
For any pair, only one between $\QF1$ and $\QF2$ is common to images used in the training set (reported in \textbf{bold}).}
	\label{tab:sensitivity_aligned}
	\footnotesize
	\subfloat[Train on $\Da_{B}^{(75)}, \, B \in \{64, 256\}$.]{\begin{tabular}{c|cc}
			\hline
			Testing $(\QF1, \QF2)$ & \textbf{$B=64$} & \textbf{$B=256$} \\ \hline
			(55, \textbf{75})                           & 0.925           & 0.982            \\
			(65, \textbf{75})                           & 0.880           & 0.981            \\
			(85, \textbf{75})                           & 0.820           & 0.952            \\ \hline
			(\textbf{60}, 78)                           & 0.900           & 0.917            \\
			(\textbf{70}, 78)                           & 0.810           & 0.907            \\
			(\textbf{60}, 80)                           & 0.860           & 0.810            \\
			(\textbf{70}, 80)                           & 0.790           & 0.800            \\ \hline
	\end{tabular}}\hfill
	\subfloat[Train on $\Da_{B}^{(85)}, \, B \in \{64, 256\}$.]{\begin{tabular}{c|cc}
			\hline
			Testing $(\QF1, \QF2)$ & \textbf{$B=64$} & \textbf{$B=256$} \\ \hline
			(55, \textbf{85})                           & 0.963           & 0.994            \\
			(65, \textbf{85})                           & 0.960           & 0.993            \\
			(75, \textbf{85})                           & 0.923           & 0.978            \\ \hline
			(\textbf{70}, 88)                           & 0.860           & 0.914            \\
			(\textbf{80}, 88)                           & 0.640           & 0.656            \\
            (\textbf{70}, 90)                           & 0.718           & 0.687            \\
			(\textbf{80}, 90)                           & 0.500           & 0.510            \\ \hline
	\end{tabular}}
\end{table}


\subsection{Non-aligned Double JPEG}

When DJPEG compression occurs with misalignment between the grids, detectors in the previous section trained on aligned data do not work anymore, getting an accuracy which is around $0.5$.
To evaluate the performance of our method for NA-DJPEG detection, we re-train the detectors in the misaligned case. In this case,  not surprisingly,  the algorithm in \cite{wang2016double} (WZ) does not work. Indeed, the features extracted by this method, i.e., the DCT histograms, are particularly distinctive only when the second compression is aligned with the first one (the typical peak and gap artifacts shows up in the DCT histograms).
%
%
Therefore, we select the well-known algorithm for NA-DJPEG detection proposed in \cite{bianchi2012detection}, denoted as BP, as additional baseline in this case.

\begin{figure}[t]
	\centering
	\subfloat[\MOD{Train on $\Dna_{256}^{ (75)}$}]{\includegraphics[height=2.2cm, trim= 0 0 3cm 0, clip=true]{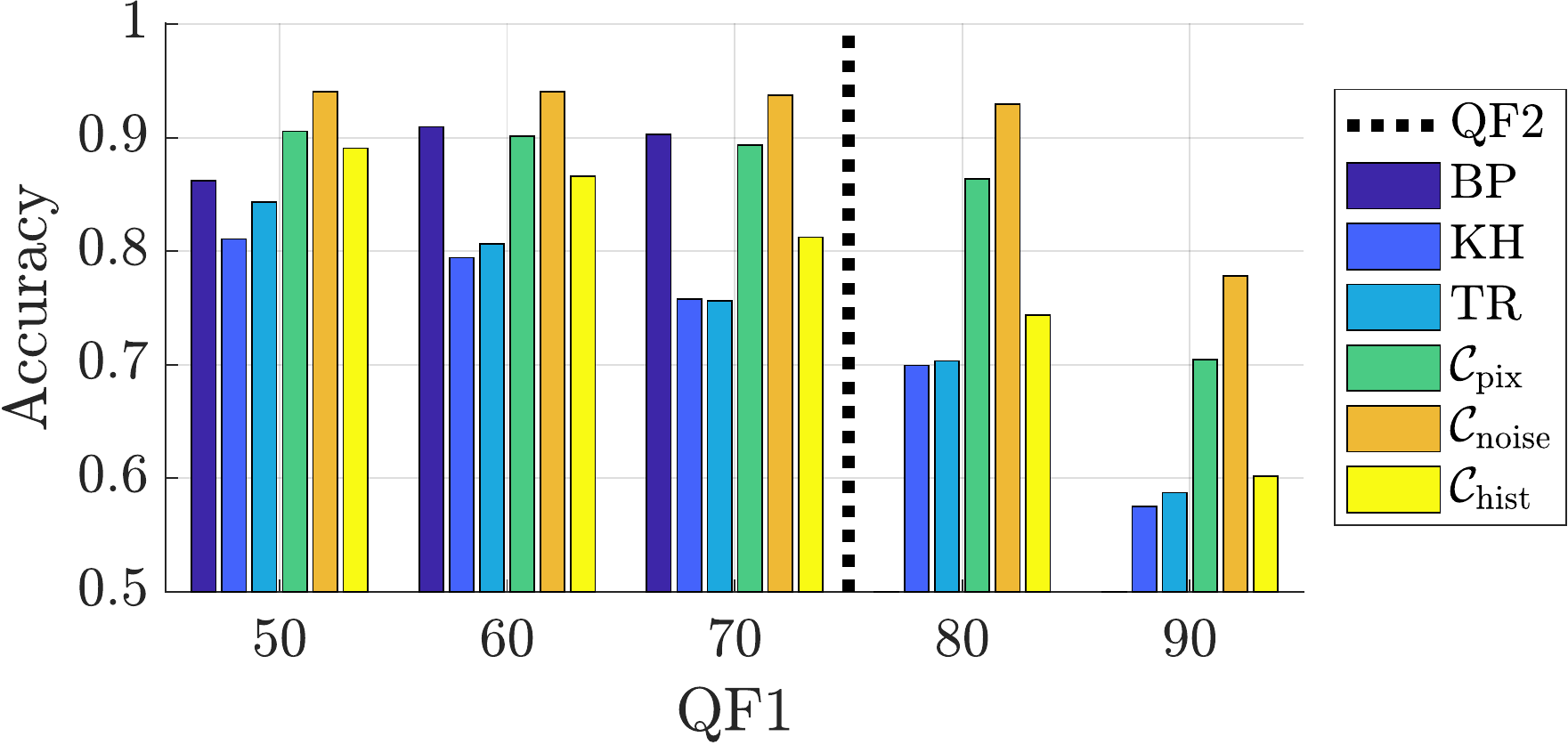}} \hfill
	\subfloat[\MOD{Train on $\Dna_{256}^{ (85)}$}]{\includegraphics[height=2.2cm, trim= 0.8cm 0 3cm 0, clip=true]{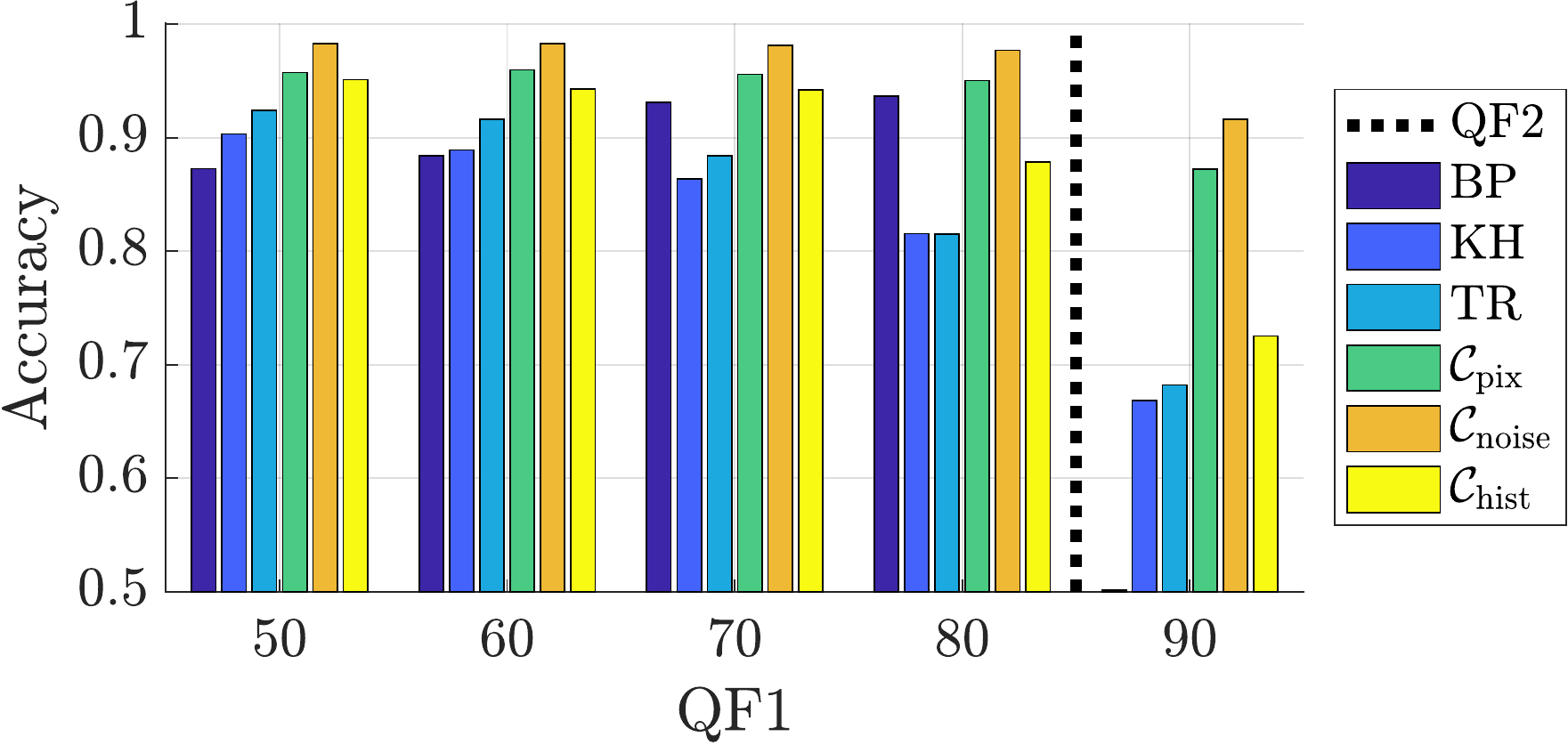}} \hfill
	\subfloat[\MOD{Train on $\Dna_{256}^{ (95)}$}]{\includegraphics[height=2.2cm, trim= 0.8cm 0 0cm 0, clip=true]{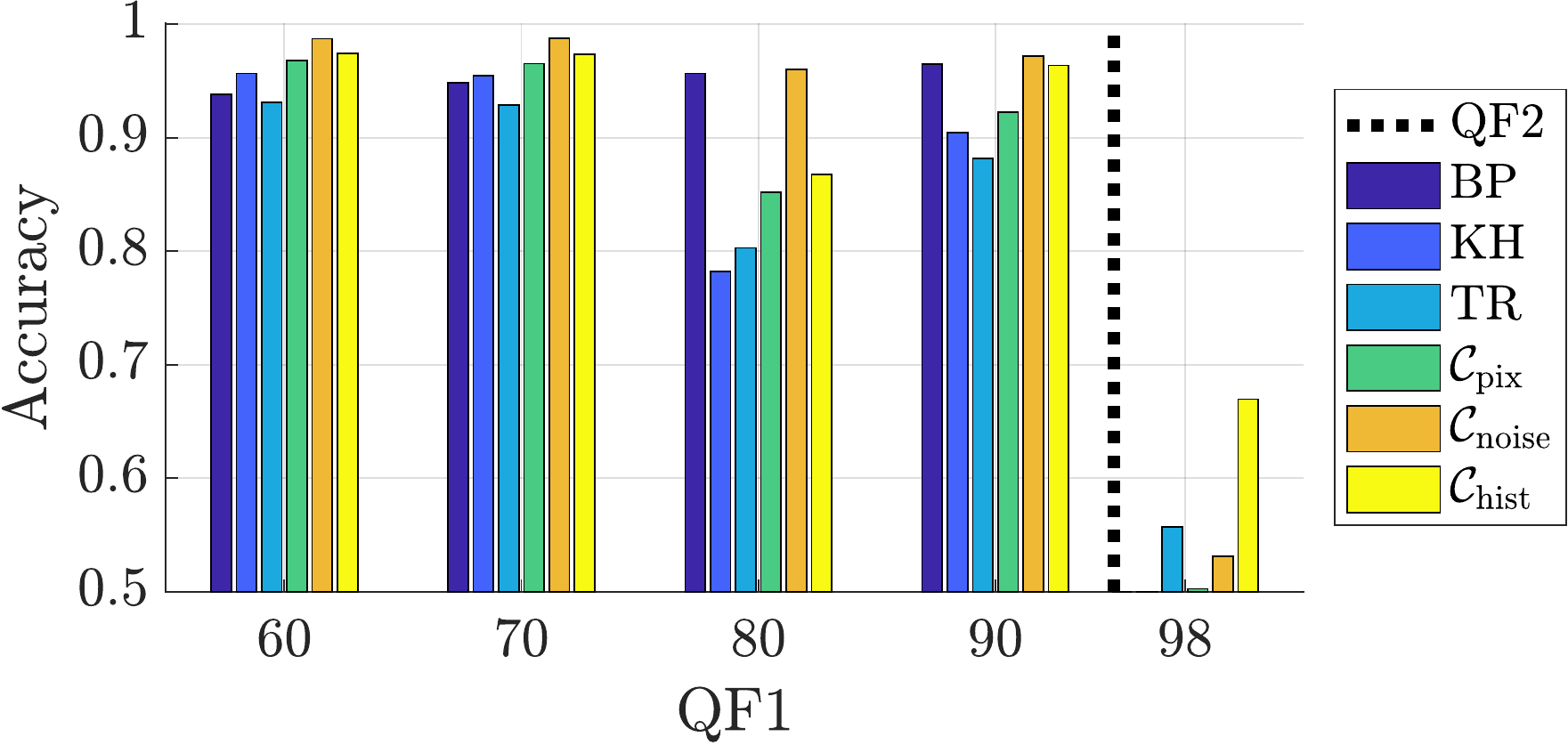}}
	
	\subfloat[\MOD{Train on $\Dna_{64}^{ (75)}$}]{\includegraphics[height=2.2cm, trim= 0 0 3cm 0, clip=true]{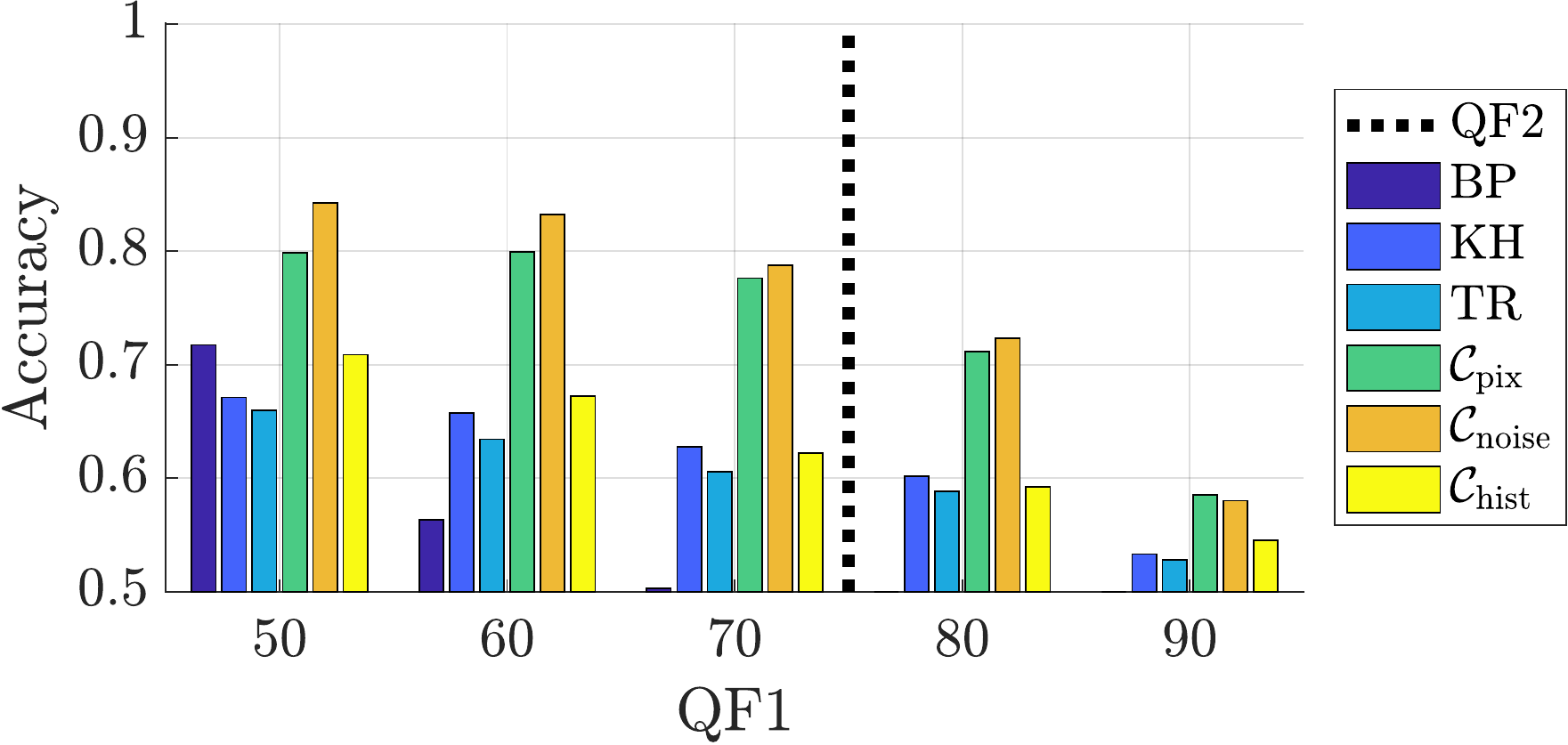}} \hfill
	\subfloat[\MOD{Train on $\Dna_{64}^{ (85)}$}]{\includegraphics[height=2.2cm, trim= 0.8cm 0 3cm 0, clip=true]{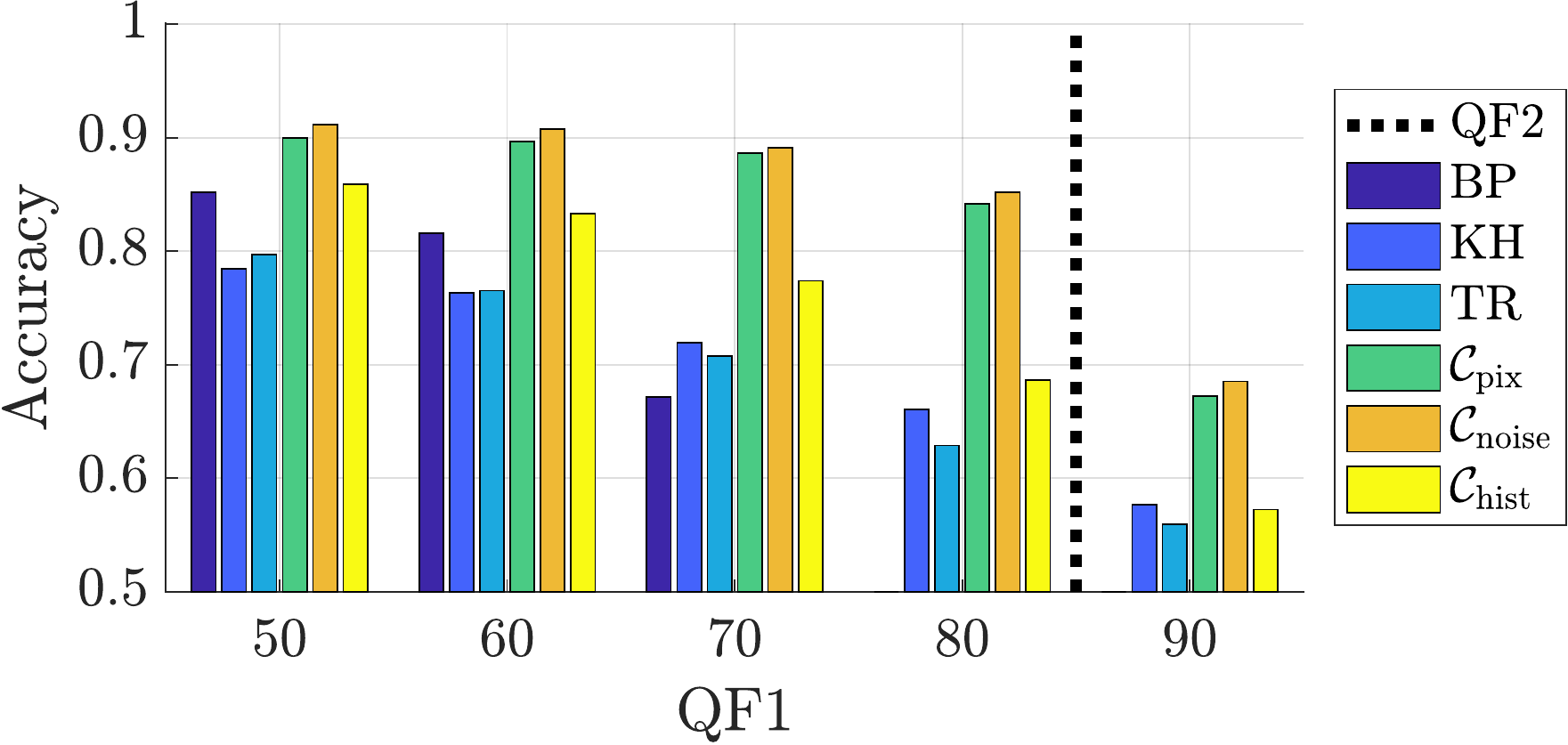}} \hfill
	\subfloat[\MOD{Train on $\Dna_{64}^{ (95)}$}]{\includegraphics[height=2.2cm, trim= 0.8cm 0 0cm 0, clip=true]{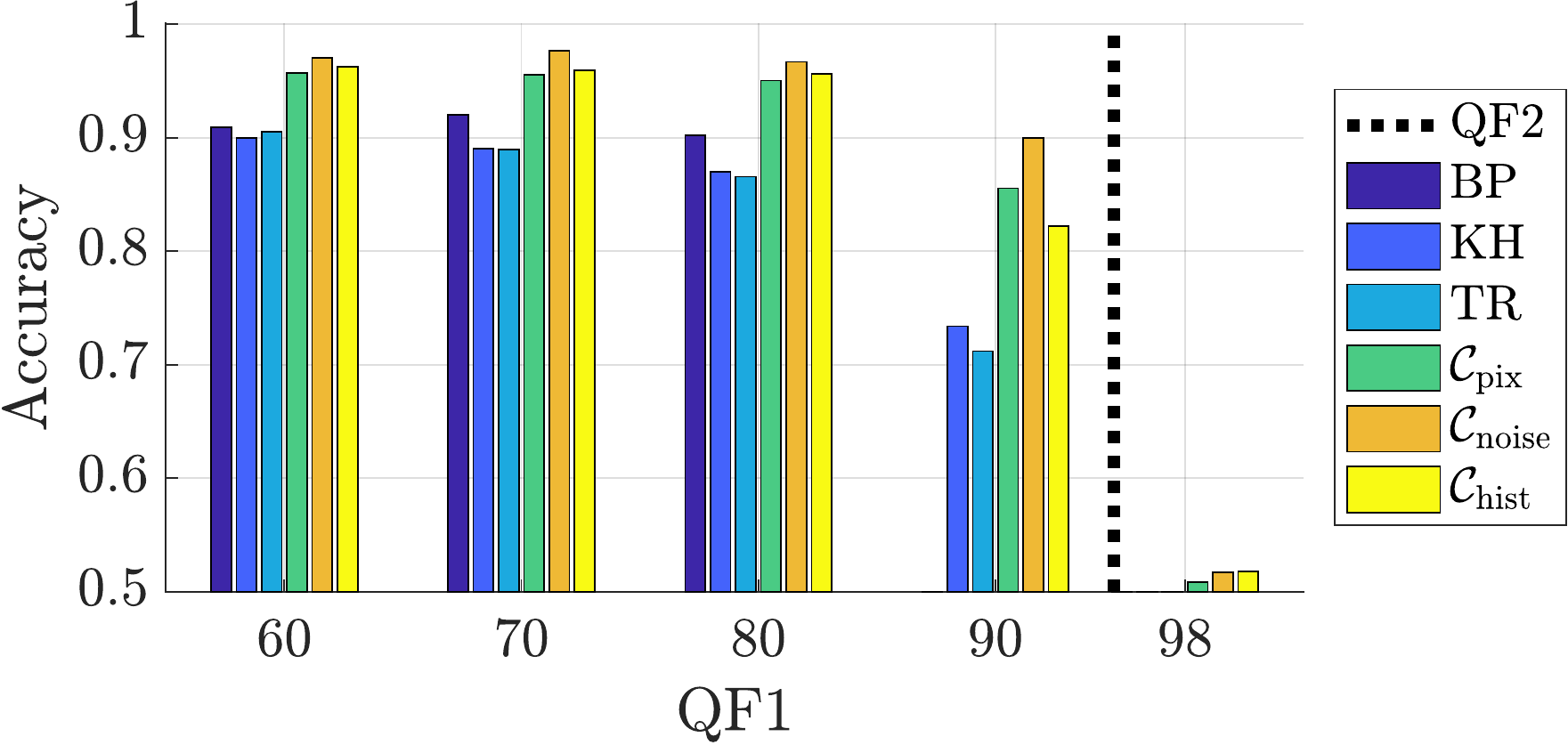}}
	\caption{\MOD{Non-aligned DJPEG compression detection accuracy against baselines BP \cite{bianchi2012detection}, KH \cite{korus2016multi}, and TR \cite{taimori2016quantization}. Dashed black line indicates the considered $\QF2$.}}
	\label{fig:acc_misaligned}
\end{figure}

Figure~\ref{fig:acc_misaligned} shows the performance of all proposed techniques and baselines for $\QF2 = 75$, $85$ and $95$ with image size $64\times 64$ and $256\times 256$.
It is known that
BP does not work when $\QF1 > \QF2$. Besides,
the accuracy significantly drops for small images, especially in the case $\QF1 \simeq \QF2$.
Concerning our methods, not surprisingly, our solution $\Chist$ shows poor performance with respect to $\Cpixel$ and $\Cnoise$.
%
Indeed, similarly to \cite{wang2016double}, the traces in the DCT domain that $\Chist$ looks at are weak in the non-aligned case.

On the other hand, CNNs designed to work in the pixel domain show good detection performance even for small images (i.e., $64\times 64$). From these results, we see that the detector based on $\Cnoise$ always outperforms state-of-the-art.
%
%

Concerning network sensitivity to QF pairs different from those in the training set, Table~\ref{tab:sensitivity_misaligned} shows the results obtained with our best method $\Cnoise$
for both $\QF1 = 75$ and $85$, and image sizes.
As for the aligned scenario, $\Cnoise$ enables good detection accuracy,
the only critical cases being those with much higher $\QF2$.
%
%
It is interesting to notice that
$\Cnoise$ is able to detect non-aligned DJPEG compression with good accuracy
also in the very challenging scenario in which $\QF1 = \QF2$.
\begin{table}[t]
	\centering
	\caption{Sensitivity of $\Cnoise$ to variations of $\QF1$ and $\QF2$ for non-aligned DJPEG detection.
Test and training images have only $\QF1$ or $\QF2$ in common (reported in \textbf{bold}).}
	\label{tab:sensitivity_misaligned}
	\footnotesize
	\subfloat[Train on $\Dna_{B}^{(75)}, \, B \in \{64, 256\}$.]{\begin{tabular}{c|cc}
			\hline
			Testing $(\QF1, \QF2)$ & \textbf{$B=64$} & \textbf{$B=256$} \\ \hline
			(55, \textbf{75})                           & 0.816          & 0.876            \\
			(65, \textbf{75})                           & 0.805           & 0.866            \\
			(75, \textbf{75})                           & 0.764         & 0.842          \\
			(85, \textbf{75})                           & 0.674          & 0.776            \\ \hline
			(\textbf{60}, 78)                           & 0.777          & 0.845            \\
			(\textbf{70}, 78)                           & 0.765           & 0.830            \\
			(\textbf{60}, 80)                           & 0.723           & 0.794            \\
			(\textbf{70}, 80)                           & 0.720           & 0.790            \\ \hline
	\end{tabular}}\hfill
\subfloat[Train on $\Dna_{B}^{(85)}, \, B \in \{64, 256\}$.]{
	\begin{tabular}{c|cc}
		\hline
		Testing $(\QF1, \QF2)$ & $B = 64$ & $B = 256$ \\ \hline
		(55, \textbf{85})            & 0.897    & 0.972     \\
		(65, \textbf{85})            & 0.878    & 0.972     \\
		(75, \textbf{85})            & 0.865    & 0.961     \\
		(85, \textbf{85})            & 0.793    & 0.954     \\ \hline
		(\textbf{70}, 88)            & 0.751    & 0.786     \\
		(\textbf{80}, 88)            & 0.738    & 0.785     \\
        (\textbf{70}, 90)            & 0.650    & 0.610     \\
		(\textbf{80}, 90)            & 0.634    & 0.600     \\ \hline
	\end{tabular}}
\end{table}


\MOD{When double compression occurs with $\QF2 = 95$ and $\QF1 > 95$, the detector fails and the images are misclassified half of the time. Experiments show that even if we train our methods to detect this specific case, the accuracy does not go above $66\%$, thus confirming that the misalignment between the $8\times 8$ compression grid tends to remove completely the traces, which in this case were already very weak in the aligned case, and then makes the detection very challenging.}

\subsection{Aligned and Misaligned Double JPEG}
\label{sec:resultsAl-NotAl}

\begin{figure}[t]
	\centering
	\subfloat[\MOD{Train on $(\Da_{64}^{(75)} \cup \Dna_{64}^{(75)})$}]{\includegraphics[width=.31\linewidth]{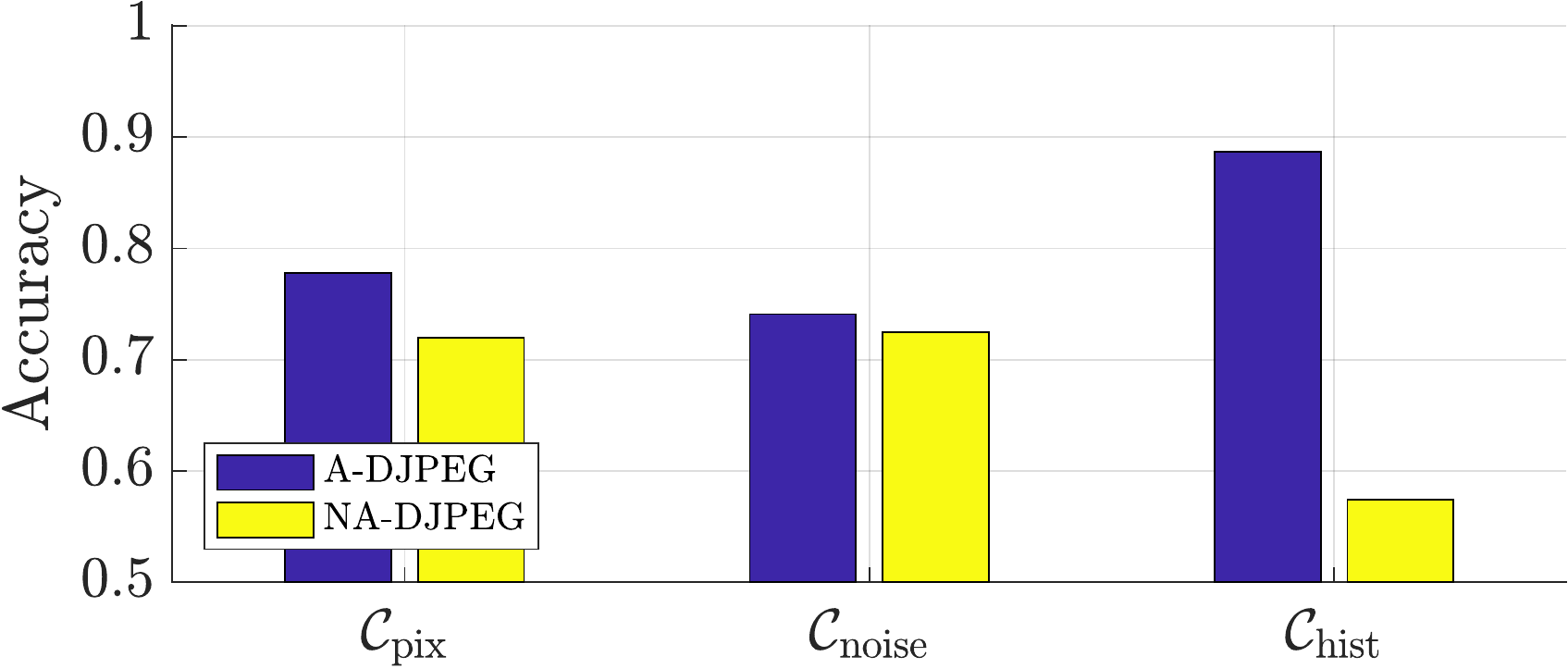}} \hfill
	\subfloat[\MOD{Train on $(\Da_{64}^{(85)} \cup \Dna_{64}^{(85)})$}]{\includegraphics[width=.31\linewidth]{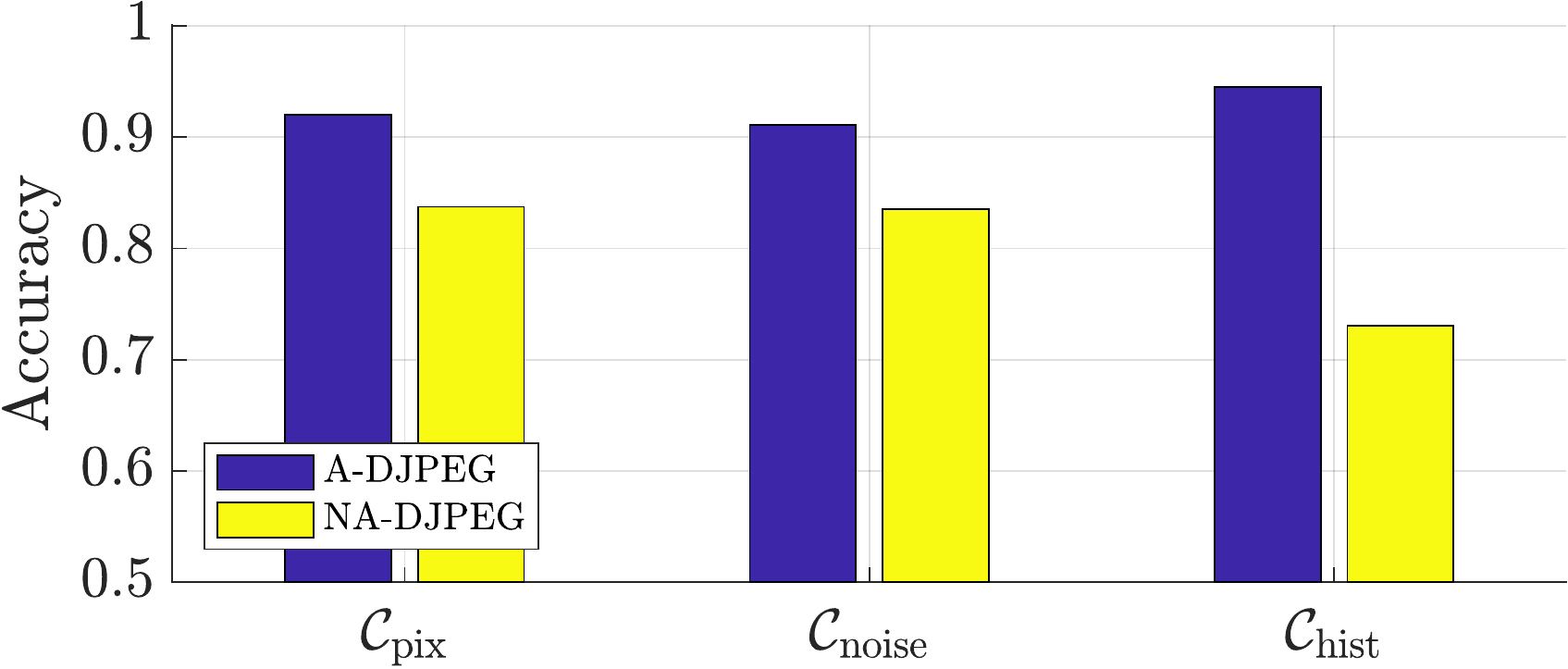}} \hfill
	\subfloat[\MOD{Train on $(\Da_{64}^{(95)} \cup \Dna_{64}^{(95)})$}]{\includegraphics[width=.31\linewidth]{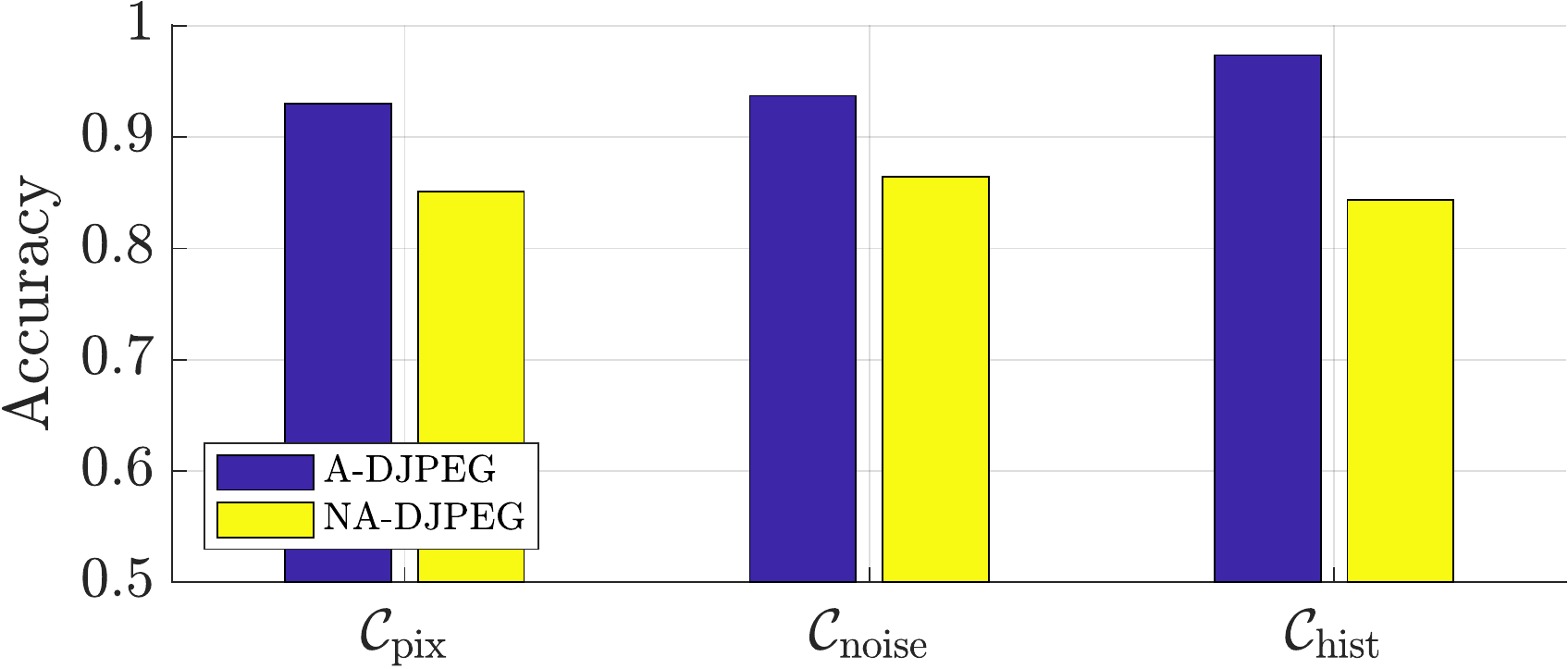}}
	\caption{\MOD{DJPEG compression detection accuracy tested separately on aligned and misaligned cases, when training is performed on a mixed dataset. Image size is 64 and $\text{QF2}=75, 85$.}}
	\label{fig:acc_total_on_all_64}
\end{figure}

Since it is usually not known a-priori whether double compression is aligned or not, it is relevant to be able to detect both A-DJPEG and NA-DJPEG. To this purpose, we trained the proposed architectures on a dataset obtained by the union of the one used for A-DJPEG, namely $\Da$, and the one used for NA-DJPEG, namely $\Dna$. For the experiments of these section, we considered the most challenging scenario with small images ($B=64$).
Figure~\ref{fig:acc_total_on_all_64} shows the performance of the CNN-based detectors in terms of average accuracy computed separately on A-DJPEG and NA-DJPEG images. The average is taken over all the $\QF$ pairs used for training. As expected from the previous analysis, $\Chist$ tends to learn better characteristics of aligned DJPEG and performs poorly in non-aligned case. Conversely, $\Cpixel$ and $\Cnoise$ are more stable solutions being able to detect with almost the same accuracy both A-DJPEG and NA-DJPEG images.

Driven by the accurate performance of $\Chist$ on A-DJPEG compression, we also investigated an alternative solution according to which the detection for the mixed case is obtained by fusing the outputs of our best CNN-based detectors for the aligned and non-aligned case,
through the use of a binary classifier. Specifically, we considered the output provided by $\Chist$ trained on A-DJPEG images, and the output of $\Cnoise$ trained in the NA-DJPEG case, as feature vector. By feeding this feature vector to a binary classifier (i.e., a random forest in our case), it is possible to further increase the final accuracy in the mixed case by up to $2\%$. However, other solutions and fusing strategies might be investigated. We leave a thorough investigation of this case to future studies.


\begin{figure}[t]
	\centering
    \includegraphics[width=1\linewidth]{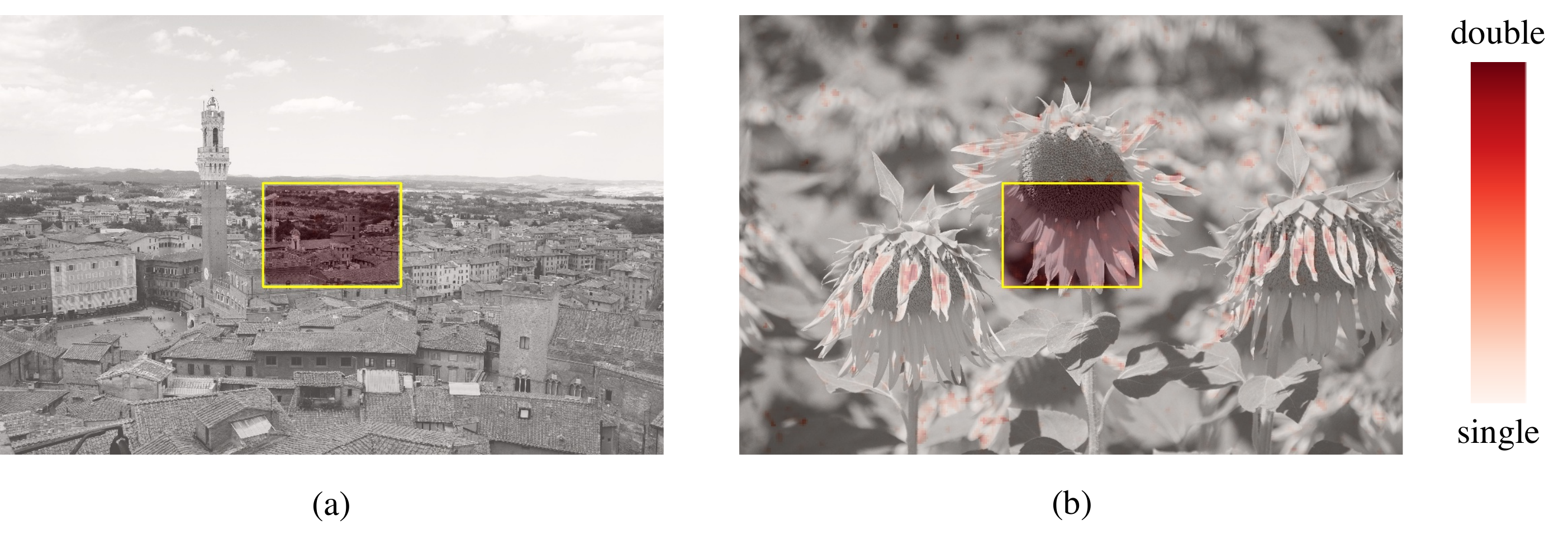}
	\caption{\MOD{A-DJPEG (a) and NA-DJPEG (b) localization example of a compressed central region with $QF2=95$ and $QF1=80$. $\Chist$ is used for the aligned case while $\Cnoise$ for the non aligned case. Actual forged region lies inside the yellow rectangle.}}
	\label{fig:tampering}
\end{figure}

\MOD{\subsection{Localization}	
Given the good performance achieved on small patches, our method can be applied on sliding windows to localize possible tampering regions in images. This can be done, e.g., by dividing the image into overlapping blocks of size $64 \times 64$ with stride $16 \times 16$. Each block is fed to the CNN (after a pre-processing step for the case of $\Cnoise$) and the softmax output is used as an estimation of the probability that the block is double compressed. Figure~\ref{fig:tampering} shows the results of double compression localization of a central region, bounded in yellow, in A-DJPEG and NA-DJPEG scenarios with $QF2=95$ and $QF1=80$, when $\Chist$ is used for the former case and $\Cnoise$ for the latter case. Both examples show that red-shaded blocks, i.e. those for which the probability of being double compressed is higher, are mainly inside the expected central region.
}




\section{Concluding remarks}\label{sec:conclusions}

In this paper we explored the use of CNNs for double JPEG compression detection problem in the case of aligned and non-aligned recompression. Specifically, three different solutions are investigated: in one of them, the CNN is based on hand-crafted features extracted from the images; in the other two, the CNN is trained directly with the images and the denoised versions, then features are self-learned by the CNN itself.
%
Results show that CNN based on hand-crafted features allow to achieve better accuracies in the case of A-DJPEG.
For the NA-DJPEG instead, the CNN based on self-learned features applied to the image noise residuals is shown to
outperform the state-of-the-art in every tested scenario.
Good performance are achieved even in the difficult cases in which the second quality factor is larger than the first and over small images,
thus paving the way to the application of the techniques to tampering localization.
Besides, CNN based on self-learned features prove very robust to deviations between training and test conditions.
%
Additionally, some preliminary experiments show the
proposed CNN-based methods can also be successfully applied to simultaneously detect an aligned or non-aligned DJPEG compression.
%

\MOD{
We designed our methods by assuming that no processing operation occurred in the middle of the two compression stages. Thought this is a common assumption to D-JPEG detection approaches in the literature,
in real applications, some intermediate processing might be applied.
%
In view of this, we made some preliminary tests  to check if and at which extent the D-JPEG detector is robust to basic processing operations\footnote{\MOD{Results are referred to $\Cnoise$ trained on both aligned and misaligned D-JPEG compressed images (Section \ref{sec:resultsAl-NotAl}) with $\QF2 = 85$ and $B=256$.}}.
The tests show that good resilience is achieved on the average with respect to histogram enhancement operations (accuracy around 85\%) and cropping (80\%), which just introduces a $8\times8$ grid desynchronization as a main effect. On the other side, the performance with respect to filtering operation are poor (62\% of accuracy in the case of a light blurring, performed with a $3\times 3$ Gaussian smoothing
kernel with variance $\sigma^2 =1$). The classification fails in the case of geometric transformation, e.g., resizing (around 30\%).
Future works will be devoted to study fusion techniques to make the most out of each network in a mixed aligned and non-aligned DJPEG case. Moreover, it would be interesting to extend the approach derived in \cite{BTeusipco} for SVM classifiers, exploiting the idea that robustness to heterogeneous processing and anti-forensics attacks can be recovered by training an adversary-aware version of the classifier.
}

\section*{Acknowledgment}
This material is based on research sponsored by DARPA and Air Force Research Laboratory (AFRL) under agreement number FA8750-16-2-0173. The U.S. Government is authorized to reproduce and distribute reprints for Governmental purposes notwithstanding any copyright notation thereon. The views and conclusions contained herein are those of the authors and should not be interpreted as necessarily representing the official policies or endorsements, either expressed or implied, of DARPA and Air Force Research Laboratory (AFRL) or the U.S. Government.

\section*{References}

{\footnotesize
\bibliography{biblio}}

\end{document}